\documentclass[10pt,twocolumn,showpacs,preprintnumbers,amsmath,amssymb,aps,prd,nofootinbib,superscriptaddress,longbibliography]{revtex4-1}
\usepackage{wasysym}
\usepackage[utf8]{inputenc}
\usepackage[english]{babel}
\usepackage{epsfig}
\usepackage{subfigure}
\usepackage{bm}
\usepackage{amsfonts}
\usepackage{dcolumn}
\usepackage{hyperref}

\hypersetup{colorlinks=true, linkcolor=blue, citecolor=green}

\usepackage{dcolumn}
\usepackage{bm}
\usepackage{ifpdf}
\usepackage{bm}
\usepackage{xcolor,color,graphicx,graphics}
\usepackage[OT1]{fontenc}
\usepackage{latexsym,amssymb,amsmath,amsfonts}
\usepackage{makeidx}
\usepackage{epstopdf}
\usepackage{mathrsfs}
\hypersetup{colorlinks=true, linkcolor=blue, citecolor=green}
\usepackage{enumerate}
 \usepackage{multirow}

\begin{document}

\title{Tidal dynamics and stellar disruption in charged Kalb–Ramond black holes in nonlinear electrodynamics}


	\author{Ednaldo L. B. Junior} \email{ednaldobarrosjr@gmail.com}
\affiliation{Faculdade de F\'{i}sica, Universidade Federal do Pará, Campus Universitário de Tucuruí, CEP: 68464-000, Tucuruí, Pará, Brazil}
\affiliation{Programa de P\'{o}s-Gradua\c{c}\~{a}o em F\'{i}sica, Universidade Federal do Sul e Sudeste do Par\'{a}, 68500-000, Marab\'{a}, Par\'{a}, Brazill}

\author{Herlan N. Lemos} \email{Herlanlemos@gmail.com}
\affiliation{Faculdade de F\'{i}sica, Universidade Federal do Pará, Campus Universitário de Tucuruí, CEP: 68464-000, Tucuruí, Pará, Brazil}
\affiliation{Programa de P\'{o}s-Gradua\c{c}\~{a}o em F\'{i}sica, Universidade Federal do Sul e Sudeste do Par\'{a}, 68500-000, Marab\'{a}, Par\'{a}, Brazill}

 \author{Marcos V. de S. Silva}
	\email{marcos.sousa@uva.es}
\affiliation{Department of Theoretical Physics, Atomic and Optics, Campus Miguel Delibes, \\ University of Valladolid UVA, Paseo Bel\'en, 7,
47011 - Valladolid, Spain}
     

\begin{abstract}
We investigate tidal forces, geodesic deviation, and tidal disruption in the black hole spacetime described by the Kalb–Ramond–ModMax solution, where electromagnetic nonlinearity is governed by the parameter $\gamma$ and Lorentz symmetry violation by the parameter $l$. In the canonical sector ($\alpha=1$), the radial tidal force exhibits a transition marked by a sign inversion between the horizons $r_{-}$ and $r_+$, signaling internal regimes of radial compression analogous to those of charged black holes; the parameter $l$ controls the strength and location of this transition, while $\gamma$ regulates the nonlinear electromagnetic contribution. The angular tidal force is predominantly compressive, $l$ shaping the effective geometry, and $\gamma$ acting as a damping factor. In the phantom sector ($\alpha=-1$), tidal forces and geodesic deviation diverge, indicating a tidal instability, with $l$ and $\gamma$ affecting only the magnitude of the response. We further show that $l$ shifts the relation between the horizon radius $r_+$ and the tidal disruption radius $r_{\rm Roche}$, thereby modifying the critical (Hills) mass defined by $r_{\rm Roche}=r_+$. Tidal disruption of neutron stars occurs inside the horizon for supermassive black holes, whereas Sun-like stars are disrupted outside the horizon, with $\gamma$ becoming relevant only for ultramassive black holes with masses $\sim10^{8}M_{\odot}$. Our results demonstrate that Kalb-Ramon–ModMax effects are largely suppressed for supermassive black holes, but may be relevant for intermediate-mass systems and observable tidal disruption events, offering an indirect probe of Lorentz violation and nonlinear electrodynamics in the strong-field regime.
\end{abstract}

\date{\today}

\maketitle

\section{Introduction}
The detection of gravitational waves by the LIGO/Virgo collaborations, originating from the coalescence of binary systems, has provided experimental evidence of the dynamics of compact objects in the strong-field regime \cite{LIGOScientific:2016aoc,LIGOScientific:2016vlm,LIGOScientific:2016sjg,LIGOScientific:2017vwq,LIGOScientific:2021qlt,LIGOScientific:2025rsn,LIGOScientific:2025rid}. Simultaneously, Event Horizon Telescope (EHT) observations of the shadows of M87* and Sgr A* enabled analyses of black hole (BH) structure in regions near the event horizon \cite{EventHorizonTelescope:2019dse,EventHorizonTelescope:2019uob,EventHorizonTelescope:2019jan,EventHorizonTelescope:2019ths,EventHorizonTelescope:2019pgp,EventHorizonTelescope:2019ggy,EventHorizonTelescope:2021bee,EventHorizonTelescope:2021srq,EventHorizonTelescope:2022wkp,EventHorizonTelescope:2022apq,EventHorizonTelescope:2022wok,EventHorizonTelescope:2022exc,EventHorizonTelescope:2022urf,EventHorizonTelescope:2022xqj}. These observational data set bounds that allow us to test General Relativity (GR) and motivate the study of alternative gravity theories at energy scales where Einstein's theory may show deviations \cite{Barack:2018yly}.

Black holes arise mathematically as solutions to Einstein field equations that feature an event horizon and a singularity \cite{Chandrasekhar:1985kt,Wald:1984rg}. In theoretical physics, they serve to probe the gravitational field and the interaction of fundamental fields under extreme conditions \cite{Berti:2009kk}. Just as experiments involving gravitational waves and BH shadows help assess the scope of GR, they also act as parameter filters for BH solutions \cite{Vagnozzi:2022moj,Barack:2018yly}.

Astrophysical BHs are not isolated systems; matter and fields are present in their surroundings \cite{Frauendiener:1990nao,Visser:1992qh}. Thus, one way to study the properties of a BH, or to extract information about its spacetime, is to examine those fields/matter around such astrophysical objects \cite{Macedo:2015ikq,Fonseca:2025ehf,Destounis:2025dck,Cardoso:2020nst,Cardoso:2020lxx}. An application of this approach is the study of tidal forces. Tidal forces are a manifestation of the gravitational gradient exerted by a massive body on an extended object \cite{dInverno:1992gxs}. Unlike the interaction between point masses, the tidal effect arises from the variation of gravitational strength across the object: the side closer to the source experiences a larger acceleration than the center of mass and the far side, generating a differential force that stretches along the alignment axis and compresses transversely.

In the BH context, tidal force studies are grounded in the concept of geodesic deviation \cite{dInverno:1992gxs,Crispino:2016pnv}. In these regions, spacetime curvature is so pronounced that the gravitational gradient between nearby points becomes very large, leading to spaghettification \cite{dInverno:1992gxs}. A key aspect is the relation between BH mass and the Roche limit: for stellar-mass BHs, tidal forces can disrupt an object well before it reaches the event horizon \cite{Hobson:2006se,Servin:2016sog}; for supermassive BHs, a body may cross the horizon intact due to the lower tidal intensity at the boundary \cite{Komossa:2015qya,Andre:2024bia}. This phenomenon underlies tidal disruption events, in which entire stars are torn apart as they approach the BH, producing highly energetic accretion disks \cite{Komossa:2015qya,Andre:2024bia}.

Since there are many classes of compact objects with distinct features as BHs, regular BHs, naked singularities, black bounces, wormholes, and others, different effects can arise in the tidal forces. For Schwarzschild, tidal forces diverge near the singularity \cite{dInverno:1992gxs}, making spaghettification especially severe \cite{Hobson:2006se}. For Reissner-Nordström (RN) and several other charged models, a sign change can occur in tidal forces, yielding a compression regime instead of stretching \cite{Crispino:2016pnv,Crispim:2025cql,Sharif:2018gaj,Lima:2020wcb,Uniyal:2025sdr,Sharif:2018gzj,Cordeiro:2025cfo}. Tidal forces have been studied in many other scenarios \cite{Albacete:2024qja,Vieira:2025vwe,Harko:2012ve,Arora:2023ltv,Silva:2025hwl,LimaJunior:2022gko}; since each case exhibits specific differences, it is natural to expect each model to modify the Roche limit accordingly \cite{Komossa:2015qya,Andre:2024bia,Joshi:2025nzb}.

Just as GR has its limitations, similar situations arise in other sectors, including the electromagnetic one. Maxwell electrodynamics successfully accounts for a wide range of electromagnetic phenomena, yet it is natural to regard it as an effective description, expected to receive corrections in strong-field regimes or whenever quantum effects become relevant. In this context, nonlinear electrodynamics (NED) provides a controlled extension of Maxwell theory and often emerges as an effective action: the Euler-Heisenberg theory encodes one-loop QED vacuum-polarization effects \cite{Heisenberg:1936nmg}, while the Born-Infeld model was originally proposed to soften classical divergences by introducing an upper bound on the electric field and later reappeared in string-theory settings \cite{Born:1934gh,Born:1934ji}. More recently, Bandos and collaborators \cite{Bandos:2020jsw}, introduced ModMax as a one-parameter nonlinear deformation that preserves two hallmark symmetries of Maxwell theory in four dimensions, namely conformal invariance and continuous $SO(2)$ electric--magnetic duality. In particular, ModMax reduces smoothly to Maxwell in the limit where the deformation parameter vanishes, $\gamma\to0$, while remaining nonlinear for finite values of this parameter; its formulation can be expressed in terms of the standard electromagnetic invariants and retains a traceless stress-energy tensor in $D=4$, in accordance with conformal symmetry. When coupled to gravity, NED sources lead to charged BH solutions that deviate from the RN geometry \cite{Bronnikov:2000vy} and, in certain scenarios, improve the interior behavior of the spacetime, thereby motivating the construction of regular BHs (for instance, the Bardeen model reinterpreted within NED \cite{Ayon-Beato:2000mjt,Rodrigues:2018bdc}) and the exploration of their geometric and observational properties. In this setting, NEDs have become a recurrent ingredient in BH modeling, regular or singular, and in the characterization of their physical observables \cite{Ayon-Beato:1999kuh,Ayon-Beato:1998hmi,Dymnikova:2004zc,Balart:2014cga,deSSilva:2024fmp,Rodrigues:2017yry}. Within this broader program, BH solutions in Einstein gravity coupled to ModMax provide a concrete arena to assess how electromagnetic nonlinearity impacts horizon structure, thermodynamics, and optical signatures, particularly in view of the effective dynamics experienced by photons in nonlinear media \cite{Amirabi:2020mzv,Pantig:2022gih,Guzman-Herrera:2023zsv,Karshiboev:2024xxx,Guzman-Herrera:2024fkg}. An interesting feature of this model is that the nonlinearity parameter allows larger charge values, compared to the RN case, while preserving the horizons; that is, the extremal charge is larger than in RN \cite{Flores-Alfonso:2020euz}.

Although Lorentz invariance is one of the cornerstones of the Standard Model and GR, there is broad motivation to investigate possible departures from this symmetry as a window into high-energy physics and effective theories emerging from quantum-gravity proposals \cite{Colladay:1998fq,Kostelecky:1988zi,Kostelecky:2003fs}. In many scenarios, Lorentz breaking does not need to be implemented explicitly at the level of the action; instead, it may arise spontaneously when additional fields acquire nonzero vacuum expectation values, thereby selecting preferred directions or tensorial structures in spacetime and modifying the gravitational dynamics \cite{Bluhm:2004ep,Seifert:2009gi,Bluhm:2008yt}. A particularly relevant example, inspired by string theory, involves the Kalb-Ramond (KR) field $B_{\mu\nu}$, an antisymmetric rank-two tensor whose gauge-invariant field strength
$H_{\mu\nu\lambda}=\partial_{[\mu}B_{\nu\lambda]}$
can generate effective terms leading to spontaneous Lorentz violation (LV) when a nontrivial background for $B_{\mu\nu}$ is present \cite{Kalb:1974yc}. In this broader context, a variety of BH solutions supported by LV (or Lorentz breaking) fields have been constructed in different frameworks, including extensions with bumblebee-type vector fields, displaying modifications of the causal structure \cite{Liu:2024oas,Maluf:2020kgf,Maluf:2021lwh,Belchior:2025xam,DCarvalho:2021zpf}. These constructions provide a useful way to investigate how LV may affect horizon geometry, geodesic motion, and potentially observable signatures \cite{Cordeiro:2025eox,Cordeiro:2025cfo,Junior:2024tmi,Junior:2024vdk,Junior:2024ety,Duan:2023gng,Kumar:2020hgm}.

In Ref.~\cite{Sekhmani:2025jbl}, the authors propose an electrically charged KR BH configuration sourced by ModMax NED (KR-ModMax solution). Within this framework, they investigate several geometric and physical properties of the model, including regularity conditions, the quasinormal-mode spectrum, bounds/estimates for greybody factors, and aspects of Hawking radiation. The corresponding BH is described by the line element
\begin{equation}
\begin{split}
ds^2 &=  -A(r)dt^2+A(r)^{-1}dr^2
	+ r^2(d\theta^2+\sin^2\theta d\phi^2)\,,\label{KR0}\\
A(r)&=\frac{1}{1-l}-\frac{2M}{r}+\alpha\frac{e^{-\gamma}Q^2}{(1-l)^2r^2},
\end{split}
\end{equation}
where $l$ is an additional parameter that characterizes the spontaneous Lorentz symmetry-breaking, $Q$ is the electric charge of the BH, $\alpha$ is the parameter that determines whether the field exhibits canonical behavior ($\alpha=1$) or phantom behavior ($\alpha=-1$), and $\gamma$ controls the degree of nonlinearity of the electromagnetic sector, with the Maxwell limit recovered as $\gamma\to 0$. This model was also investigated in Ref. \cite {Al-Badawi:2025ejf}, where the authors focus on its thermodynamic properties as well as on features associated with the geodesic structure, including the orbits of neutral and charged particles, light deflection, and the BH shadow radius.

This work is organized as follows: in Sec.\,\ref{secparameter}, we present the causal structure of the spacetime and discuss the constraints that its parameters must satisfy. In Sec. \ref{sec:two}, we
discuss the formalism required for the study of tidal forces (geodesic deviation equation and the tidal tensor) and applied this formalism to find the radial and the angular tidal forces to the KR-ModMax BH. Using tidal forces, we compute the Roche limit in Sec. \ref{secIV}, considering both neutron stars and Sun-like stars. Geodesic deviation, i.e., the stretching or compression experienced by bodies due to tidal forces, is discussed in Sec. \ref{secV}. Finally, our concluding remarks and final conclusions are presented in Sec. \ref{Sec:Conclusion}.

\section{Parameter constraints of the solution}\label{secparameter}
The line element \eqref{KR0} has a causal structure similar to ModMax without magnetic charge, with modifications now provided by the LV parameter, given by 
\begin{equation}
r_{\pm}=M(1-l)\left[1\pm\sqrt{1-\frac{\alpha e^{-\gamma}Q^2}{M^2(1-l)^3}}\right], \label{r+-}
\end{equation}  
with $r_+$ being the event horizon and $r_{-}$ the Cauchy one. Furthermore, for $\alpha=1$, there are horizons if
\begin{equation}
0\leq Q \leq M e^{\gamma/2}(1-l)^{3/2} \,,
\end{equation}
and the critical charge is defined as $Q_{crit}/M=(1-l)^{3/2}e^{\gamma/2}$. The bounds imposed in \cite{Duan:2023gng} on the parameter $l$, using the inference of the shadow radius of Sgr~A* by the EHT Collaboration, show that
\begin{eqnarray}
-4.59\times 10^{-3} \leq l \leq 1.24\times 10^{-1}.
\end{eqnarray}
For illustrative purposes, we adopt larger positive and negative values of $l$ ($l=-0.1$ and $l=0.1$) which allows a clearer graphical visualization of the differences between the RN solution, the electrically charged KR configuration, and the usual ModMax. This choice is made solely for clarity and does not affect the physical conclusions. Thus, if $l<0$ with $\gamma>0$, the condition $(1-l)^{3/2}>1$ allows $Q/M>1$ due to charge screening by $\gamma$ and the existence of a horizon. The same occurs for $l>0$, provided $l<1$, with an attenuation of the horizon. The condition $\gamma>0$ also appears in other works when imposing unitarity and causality in ModMax theory. In fact, to ensure causality and unitarity, $\gamma$ must satisfy
$0<\gamma<\tanh^{-1}(\sqrt{2}/2\bigr)$ \cite{Amirabi:2020mzv,Bandos:2020jsw}. In \cite{Guzman-Herrera:2023zsv}, the authors also use EHT data for Sgr A* to constrain the BH charge. One of the resulting bounds is
$Q e^{-\gamma/2}\le 0.8M$.

\section{TIDAL FORCES}\label{sec:two}
The tidal effects acting on a freely falling particle in the gravitational field of a BH are characterized by the spatial components of the geodesic deviation vector $\eta^\mu$. This vector represents the infinitesimal separation between neighboring geodesics and is determined through the geodesic deviation equation \cite{dInverno:1992gxs}
\begin{equation}
\frac{D^2\eta^\mu}{D\tau^2}+R^{\mu}_{\,\,\nu\alpha\beta}v^\nu\eta^\alpha v^\beta = 0\,,
\end{equation}
where $R^{\mu}_{\,\,\nu\alpha\beta}$ is the Riemann tensor and $v^\beta$ is the unit vector tangent to the geodesic. When a test body moving along a geodesic in spacetime, its constituent points trace neighboring geodesics rather than a single identical trajectory. The resulting relative accelerations between these adjacent worldlines give rise to tidal effects, manifested as stretching and squeezing of the body. These tidal effects are most conveniently described by introducing a set of orthonormal basis vectors (tetrads) adapted to the observer's frame. Such a construction defines a local inertial reference frame attached to the object at each point along its geodesic path. The components of such tetrads $e^\mu_{\hat{a}}=(e^\mu_{\hat{0}}, e^\mu_{\hat{1}}, e^\mu_{\hat{2}}, e^\mu_{\hat{3}})$ are found in our case as
\begin{eqnarray}
e^\mu_{\hat{0}} &=& \left(\frac{E}{A(r)}, -\sqrt{E^2-A(r)}, 0, 0\right)\;,\nonumber\\ e^\mu_{\hat{1}}&=&\left(-\frac{\sqrt{E^2-A(r)}}{A(r)}, E, 0, 0\right)\,,\nonumber\\
e^\mu_{\hat{2}}&=&\left(0, 0, \frac{1}{r}, 0 \right)\,,\,\,\, e^\mu_{\hat{3}}=\left(0, 0, 0, \frac{1}{r\sin\theta}\right)\,,
\end{eqnarray}
satisfying the orthonormality condition $e^\mu_{\hat{a}}e_{\mu\hat{b}}=\eta_{\hat{a}\hat{b}}$ where $ \eta_{\hat{a}\hat{b}}={\rm diag}\left(-1, 1, 1, 1\right)$ is the Minkowski metric and $e^\mu_{\hat{0}}=v^\mu$ is the four-velocity. The geodesic deviation vector can be expanded into $\eta^\mu=e^\mu_{\hat{a}}\eta^{\hat{a}}$ and fix the time component as $\eta^{\hat{0}}=0$.

In this formalism, the components of the Riemann curvature tensor are expressed with respect to the tetrad basis according to
\begin{eqnarray}
R^{\hat{a}}_{\,\,\hat{b}\hat{c}\hat{d}}=e^{\hat{a}}_{\,\,\mu}e^\nu_{\,\,\hat{b}}e^\rho_{\,\,\hat{c}}e^\sigma_{\,\,\hat{d}}R^\mu_{\,\,\nu\rho\sigma}\,,
\end{eqnarray} 
whose nonzero components are
\begin{eqnarray}
&& R^{\hat{0}}_{\,\,\hat{1}\hat{0}\hat{1}}=-\frac{A''(r)}{2}\,,\label{R0101}\\
&& R^{\hat{0}}_{\,\,\hat{2}\hat{0}\hat{2}}=R^{\hat{0}}_{\,\,\hat{3}\hat{0}\hat{3}}=R^{\hat{1}}_{\,\,\hat{2}\hat{1}\hat{2}}=R^{\hat{1}}_{\,\,\hat{3}\hat{1}\hat{3}}=-\frac{A'(r)}{2r}\,,\label{R0202}\\
&&R^{\hat{2}}_{\,\,\hat{3}\hat{2}\hat{3}}=\frac{1-A(r)}{r^2}\,.\label{R2323}
\end{eqnarray}
Since the tetrad components $e^\mu_{\,\,\hat{a}}$ are parallel transported along the geodesic, Eqs.~\eqref{R0101}-\eqref{R2323} directly lead to the tidal force equations governing the motion of a neutral freely falling body in the vicinity of the BH, which are
\begin{eqnarray}
&&\frac{D^2\eta^{\hat{1}}}{D\tau ^2}=-\frac{A''(r)}{2}\eta^{\hat{1}}\,,\label{radial1}\\
&&\frac{D^2\eta^{\hat{i}}}{D\tau ^2}=-\frac{A'(r)}{2r}\eta^{\hat{i}}\,,\label{angulari}
\end{eqnarray}
where $\hat{1}$ is the radial component and $i=2, 3$ are the angular components $(\hat{\theta},\hat{\phi})$ of the tidal force in the local frame. 

In what follows, we analyze the tidal forces acting on a neutral test body undergoing free fall in the vicinity of the BH governed by the metric \eqref{KR0}, corresponding to the KR ModMax solution.

\subsection{Radial tidal forces}

\begin{figure*}[ht!]
\centering
\subfigure[\,\,$Q = 0.6$ and $\gamma=0.5$] 
{\label{Fradiala}\includegraphics[width=8.5cm]{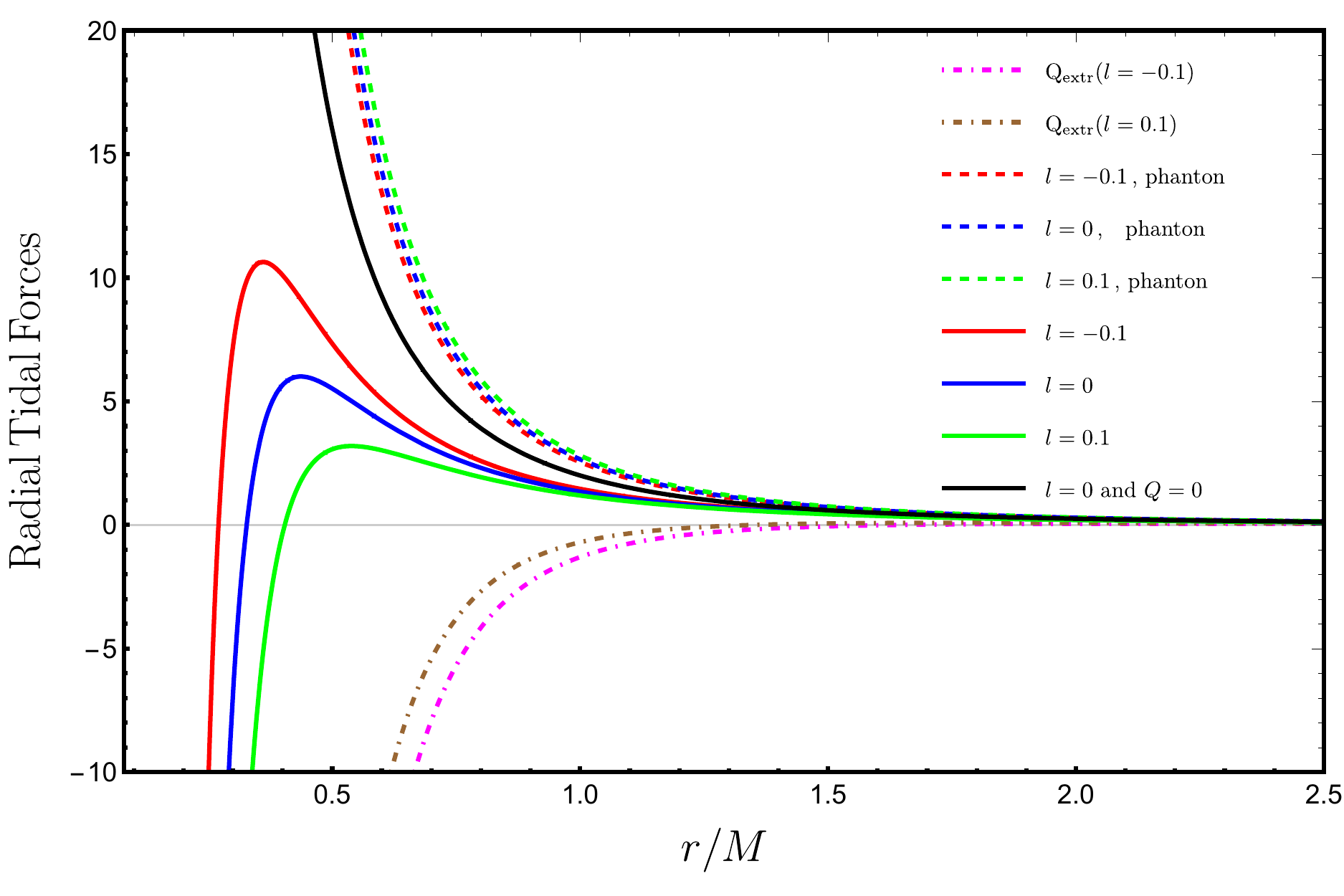} }
\hspace{0.1cm}
\subfigure[\,\, $l=1.24\times 10^{-1}$] 
{\label{Fradialb}\includegraphics[width=8.5cm]{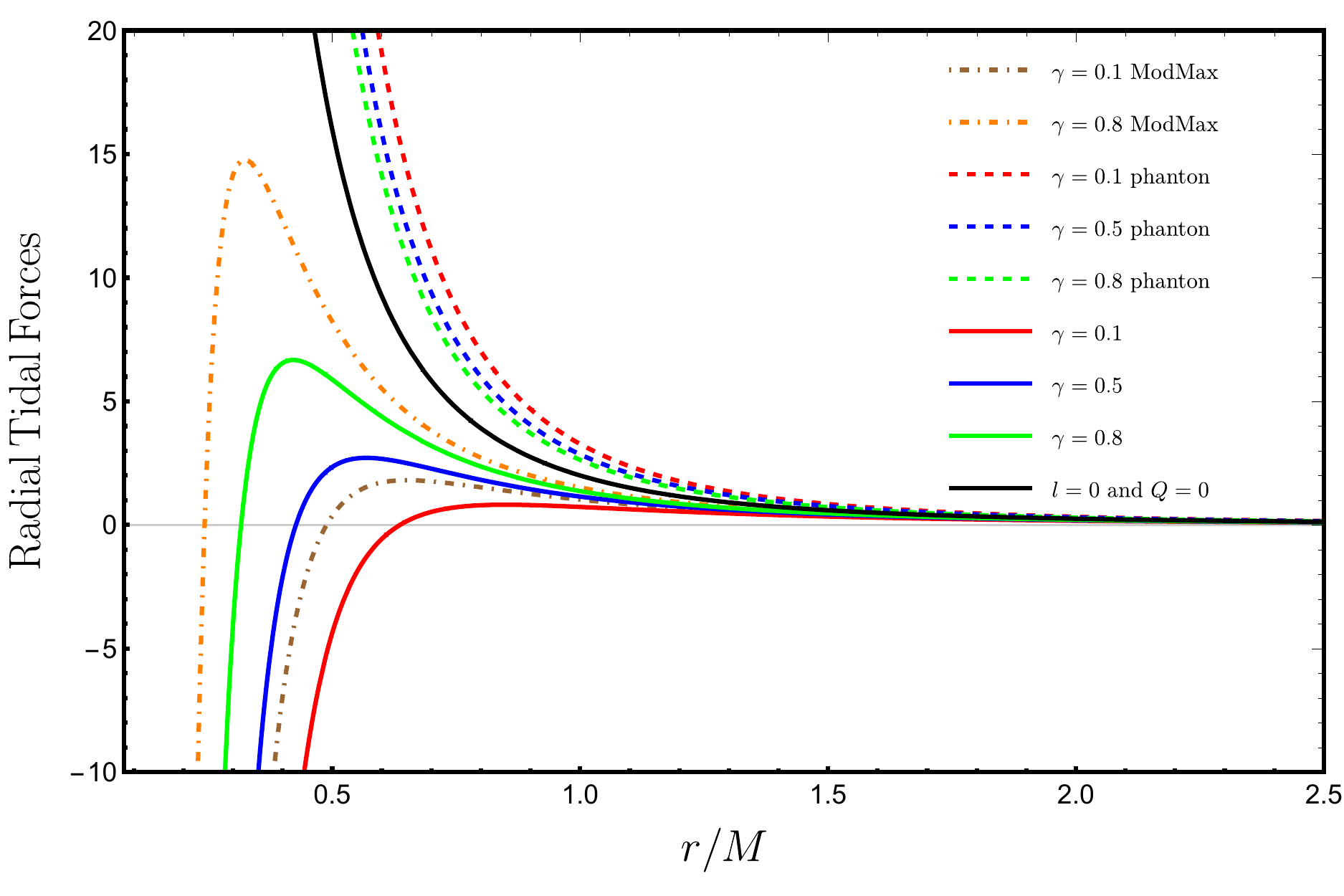} }
\caption{(a) Radial tidal forces with $Q=0.6$, $\gamma=0.5$  and different $l$.  (b) Radial tidal forces with $l=1.24\times 10^{-1}$ for different $\gamma$ and $Q=0.6$. Here we have not used the negative values of $l$ because within the range obtained in \cite{Duan:2023gng} its effect is very small, and its effects are not evident.}\label{Fig3}
\end{figure*}  

Applying Eq.~\eqref{radial1} to the metric \eqref{KR0}, we obtain the expression for the radial tidal force,
\begin{eqnarray}
\frac{D^2\eta^{\hat{1}}}{D\tau ^2}=\frac{1}{2} \left(\frac{4 M}{r^3}-\frac{6 \alpha  e^{-\gamma}Q^2}{(1-l)^2 r^4}\right)\eta^{\hat{1}}\,,\label{Fradial}
\end{eqnarray}
which recovers the charged KR case in the limit $\gamma\rightarrow 0$ \cite{Cordeiro:2025cfo}, reproduces the RN result in the limit $l\rightarrow 0$ and $\gamma\rightarrow 0$ \cite{Crispino:2016pnv} and reduces to the Schwarzschild case when $Q=0$. Unlike the Schwarzschild solution \cite{dInverno:1992gxs}, when $\alpha=1$, the radial tidal force can vanish at a specific radius $r=R^{\rm rad}_0$, which is determined from Eq.~\eqref{Fradial} as
\begin{eqnarray}
R^{\rm rad}_0=\frac{3 e^{-\gamma}Q^2}{2 (1-l)^2 M}\,,\label{radial0}
\end{eqnarray}
while it takes its maximum at
\begin{eqnarray}
R^{\rm rad}_{\rm max}=\frac{2 e^{-\gamma} Q^2}{(1-l)^2 M}\,.\label{radialmax}
\end{eqnarray}
For the phantom case, both $R^{\rm rad}_{0}$ and $R^{\rm rad}_{\rm max}$ take negative values and do not belong to the physical domain for any $l$, $M$, $Q$, and $\gamma$ within the allowed ranges.

The radial tidal force given by Eq.~\eqref{Fradial} is plotted in Fig.~\ref{Fradiala} for fixed values of $\gamma$ and $Q$, and different values of $l$. In the canonical sector ($\alpha=1$), the behavior of the radial tidal force resembles that of the KR-RN solution investigated in \cite{Cordeiro:2025cfo}, indicating the existence of a tidal transition region in which the extended particles cease to be rapidly stretched and instead become compressed. For $l<0$ (red curve), the tidal force reaches a higher intensity compared to the usual ModMax case (blue curve), while it is weaker for $l>0$ (green curve). For suitable parameter choices, the point $r=R^{\rm rad}_{0}$ lies between $r_-$ and $r_+$ and varies according to the parameters $\gamma$ and $l$.  
In the phantom sector ($\alpha=-1$), the dashed curves show a tidal force that does not vanish at any physical radius and closely resembles the Schwarzschild case (black curve), growing monotonically as $r$ decreases, with its profile being shifted by the values of $l$. The profiles corresponding to the extremal charge are also shown by the cyan and magenta dot–dashed curves for the chosen parameters.  
  
 In Fig.\,\ref{Fradialb} we plot the radial tidal force for $l=1.24\times 10^{-1}$ and different $\gamma$ for fixed values of $Q$. The ModMax parameter significantly modifies the radial tidal force. In the canonical sector ($\alpha=1$), represented by the solid red, blue, and green curves, as $\gamma$ increases the transition from radial stretching to compression becomes more pronounced as the particle approaches the singularity, with a corresponding shift in the inversion and zero points of the force. This effect is more substantial in the usual ModMax case ($l=0$ and $\alpha=1$), as indicated by the brown and orange dot–dashed curves. In the phantom sector $(\alpha=-1)$, the behavior is similar to the previous case, with no tidal transition region present.

\subsection{Angular Tidal Forces}

\begin{figure*}[ht!]
\centering
\subfigure[\,\,$Q = 0.6$ and $\gamma=0.5$] 
{\label{Fangulara}\includegraphics[width=8.5cm]{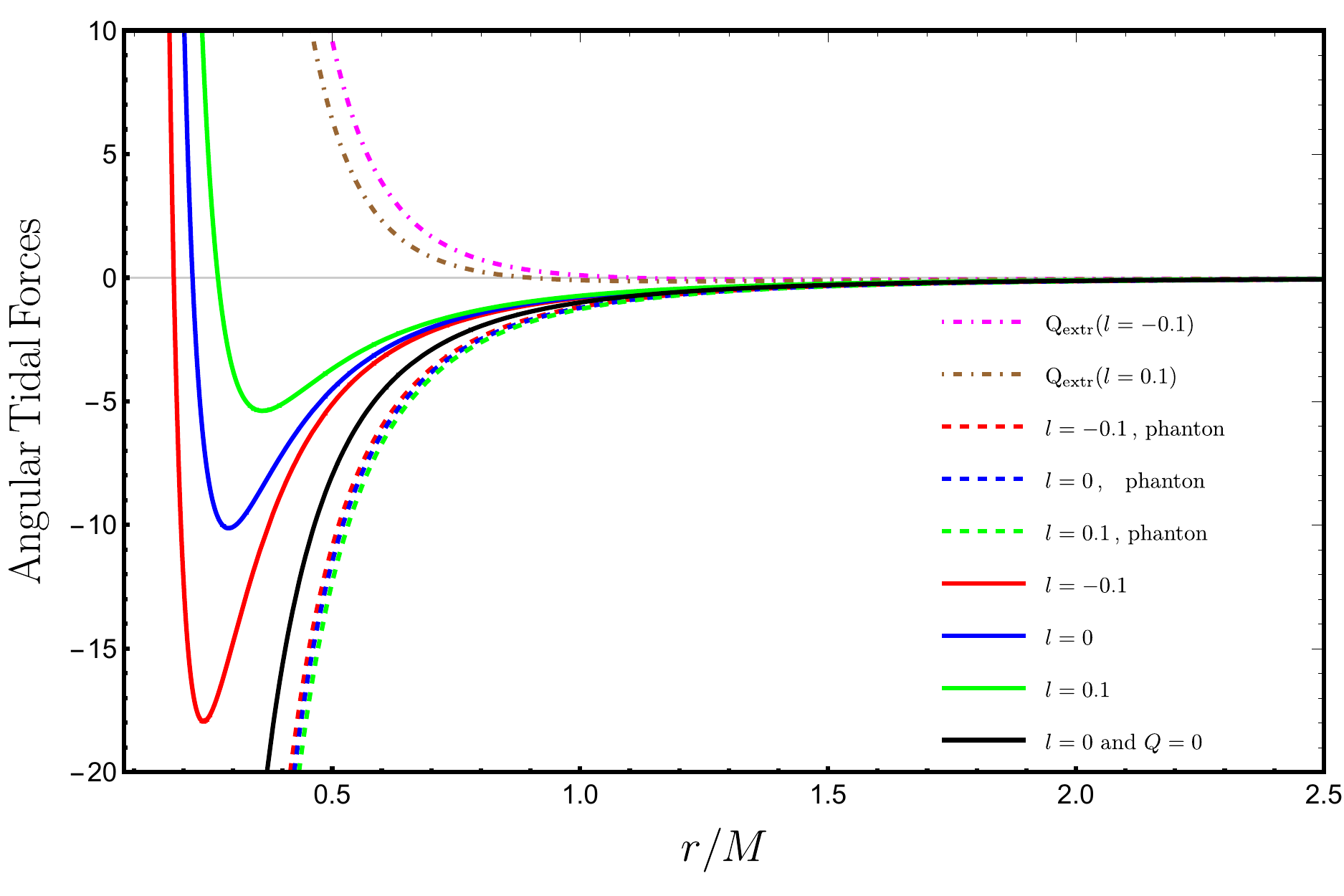} }
\hspace{0.1cm}
\subfigure[ \,\,$l=1.24\times 10^{-1}$] 
{\label{Fangularb}\includegraphics[width=8.5cm]{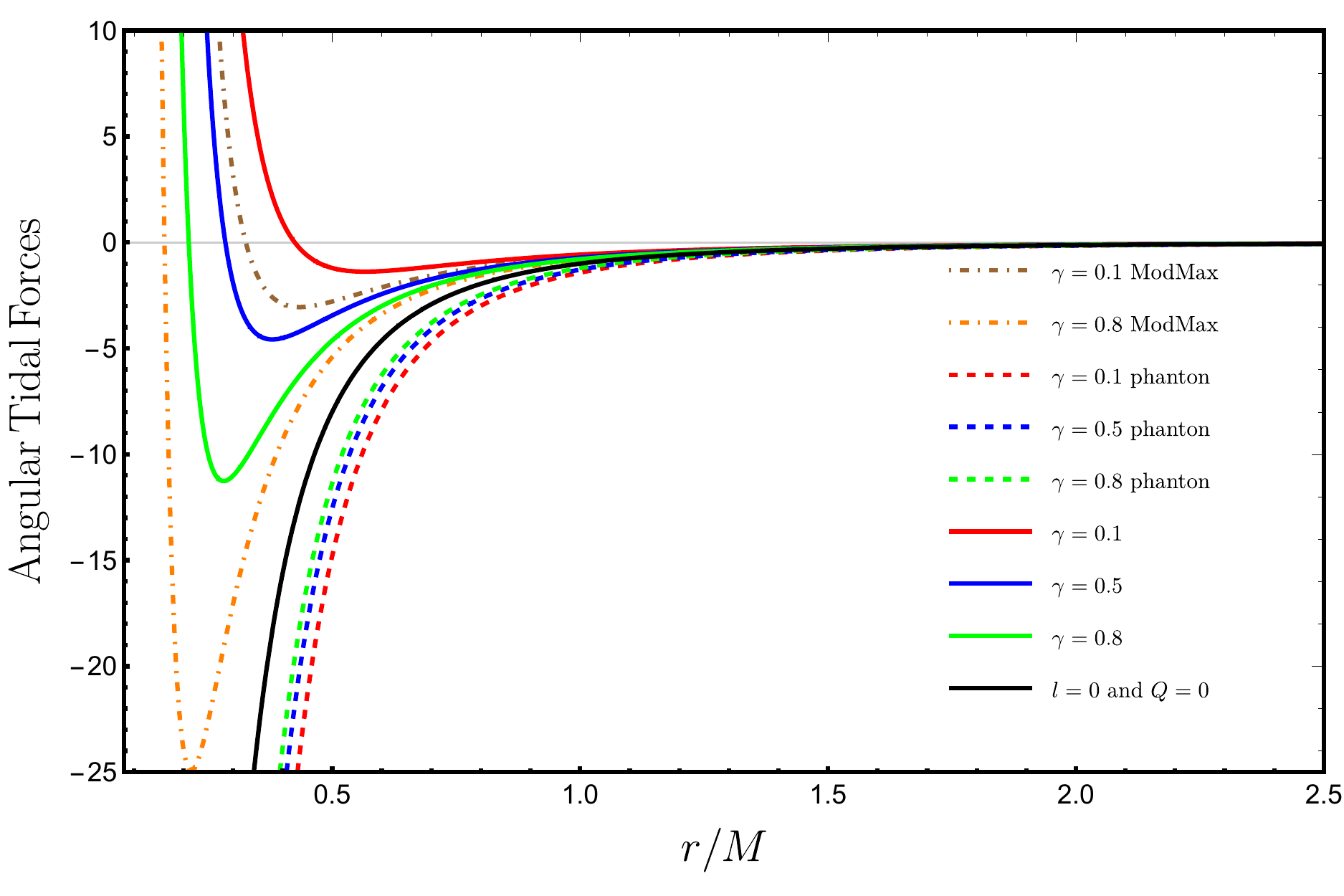} }
\caption{(a) Angular tidal forces with $Q=0.6$ and different $l$. (b) Angular tidal forces with $l=1.24\times 10^{-1}$ for different $\gamma$ and $Q=0.6$. Note that we did not used negative values of $l$ because their effects can be neglected within the range obtained in \cite{Duan:2023gng}.}\label{Fig4}
\end{figure*}

Using Eqs.\,\eqref{KR0} and \eqref{angulari}, the angular tidal force in the KR-ModMax BH is
\begin{equation}
\frac{D^2\eta^{\hat{i}}}{D\tau ^2}=\frac{1}{r^4}\left(\frac{\alpha e^{-\gamma}Q^2}{(1-l)^2}-M r\right)\eta^{\hat{i}}\,, \label{Fangular}
\end{equation} 
which recovers the charged KR case in the limit $\gamma\rightarrow 0$ \cite{Cordeiro:2025cfo}, reproduces the RN result in the limit $l\rightarrow 0$ and $\gamma\rightarrow 0$ \cite{Crispino:2016pnv} and reduces to the Schwarzschild case when $Q=0$. For $\alpha=1$, the angular tidal force is zero at 
\begin{equation}
R^{\rm ang}_0=\frac{e^{-\gamma}Q^2}{(1-l)^2 M}\,,\label{angular0}
\end{equation}
and minimum at
\begin{equation}
R^{\rm ang}_{\rm min}=\frac{4e^{-\gamma} Q^2}{3 (1-l)^2 M}\,. \label{angularmax}
\end{equation} 
For the phantom case, both $R^{\rm ang}_0$ and $R^{\rm ang}_{\rm min}$ take negative values and do not belong to the physical domain, for any $l$, $M$, $Q$, and $\gamma$ within the allowed ranges.

The curves corresponding to the angular component of the tidal force are shown in Fig.\,\ref{Fangulara}. In the canonical sector (solid red, blue, and green curves), the angular tidal force is negative over a large part of the external region, indicating angular compression, analogously to the RN case \cite{Crispino:2016pnv} and the KR–RN solution \cite{Cordeiro:2025cfo}. For $l<0$ (red curve), the angular compression is enhanced, making the force more negative. For $l=0$ (blue curve), the behavior is intermediate, while for $l>0$ (green curve) the angular compression is weakened. This shows that the parameter $l$ controls the profile of the tidal force.

The black curve corresponds to the Schwarzschild case ($Q=0$ and $l=0$), for which the angular tidal force is always negative and grows as $r\rightarrow 0$, leading to the typical transverse compression associated with spaghettification. In the phantom sector (dashed curves), the qualitative behavior is similar to the Schwarzschild case, but shifted by the values of $l$ and with no sign inversion. The magenta and brown dot–dashed curves correspond to the extremal charge for different values of $l$, in this case, the angular profiles are strongly modified by the nonlinearity of the electromagnetic field, indicating that the angular tidal force is highly sensitive to extremality.

Figure\,\ref{Fangularb} displays the behavior of the angular component of the tidal force for fixed $l=1.24\times 10^{-1}$ and $Q=0.6$ and different values of $\gamma$. In the canonical sector (solid red, blue, and green curves), the angular tidal force is negative over most of the external region, indicating angular compression. As $\gamma$ increases, the term $e^{-\gamma}Q^2$ decreases, reducing the nonlinear electromagnetic contribution to the angular tidal force component. This leads to a progressively smaller magnitude of angular compressive force, with the force minimum approaching the Schwarzschild case. Therefore, the parameter $\gamma$ acts as a damping and shaping factor, weakening the ModMax effects and smoothing the angular tidal profile, in contrast to $l$, which controls the effective geometry and the location of the extrema.

By comparison with Fig.\,\ref{Fangulara} no significant shift of $R^{\rm ang}_{\rm min}$ is observed. The parameters analyzed here thus play complementary roles in determining the structure of the angular tidal force.

In the phantom sector (dashed curves), the angular tidal force is positive, indicating angular stretching, as in the Schwarzschild case (black curve), but the angular behavior is identical for all values of $\gamma$. The phantom sector is therefore qualitatively insensitive to variations in $\gamma$. The brown and orange curves correspond to the usual ModMax case ($l=0$) with different values of $\gamma$. ModMax nonlinearity is essential to modify the angular component of the tidal force.

When the charge parameter is set to $Q=0$ the metric in Eq. \eqref{KR0} reduces to the neutral KR background \cite{Yang:2023wtu}. In this limit, the radial and angular tidal forces recover exactly the Schwarzschild-like behavior, since both components depend solely on derivatives of the metric function $A(r)$, as shown in Eqs. \eqref{radial1} and \eqref{angulari}. Consequently, the LV parameter does not influence the free fall dynamics in the uncharged configuration.

The situation changes once the ModMax electromagnetic sector is considered. For $Q\neq0$, nonlinear electromagnetic contributions explicitly enter the tidal force components, leading to deviations from the Schwarzschild profile. In this case, the tidal behavior of the KR-ModMax solution also departs from that of the standard RN spacetime, as the ModMax nonlinearity modifies both the magnitude and the radial dependence of the electromagnetic contribution to the tidal forces. As a consequence, the combined KR-ModMax system exhibits tidal features that are distinct from those of the uncharged KR, Schwarzschild, and RN solutions, revealing new effects associated with NED in the strong field regime.

The existence of $r=R^{\rm rad, ang}_{0}$ between the inner and outer horizons indicates a nontrivial internal tidal structure, analogous to that of charged BHs, although such a transition is hidden from external observers. The presence of such a tidal transition region may modify tidal disruption processes, since extended objects can experience a regime of weaker or qualitatively different forces compared to those expected for Schwarzschild or standard RN BHs. In the following section, we briefly investigate the tidal disruption process associated with the KR-ModMax BH.

\section{TIDAL DISRUPTION RADIUS IN KR-MODMAX BH}\label{secIV}

In the KR-ModMax solution, the introduction of the parameter $l$, associated with LV, leads to nontrivial modifications of the fundamental geometric scales of spacetime. In particular, the event horizon depends explicitly on the sign and magnitude of $l$. For $l>0$ these characteristic scales are reduced when compared with the standard Schwarzschild, RN, and ModMax cases. In contrast, for $l<0$, the horizon becomes larger than in the corresponding conventional scenarios. 

These geometric modifications have direct consequences for the dynamics of extended objects in the vicinity of the gravitational center. The contraction of the horizon allows stars to approach closer to the center before being captured, thereby enhancing the relevance of tidal forces and favoring scenarios in which tidal disruption occurs outside the horizon. Conversely, the expansion of the orbital scales tends to trigger gravitational capture at larger radii, limiting the maximum proximity that stars can reach without crossing the horizon. In this context, the Roche limit provides a direct criterion to distinguish between capture and tidal disruption regimes, allowing one to assess how LV reshapes the radial hierarchy among the horizon and the tidal disruption radius \cite{Servin:2016sog,Andre:2024bia}.

The tidal disruption radius is typically estimated by equating the radial tidal force produced by the central object to the star's self-gravitational binding at its surface. Within this standard approximation, tidal disruption is assumed to occur when the tidal force across the stellar radius matches the star's surface gravity. Accordingly, the Roche limit is obtained by equating the radial tidal acceleration to the star's surface gravity; using \eqref{radial1}, we then obtain \cite{Servin:2016sog,Andre:2024bia}
\begin{equation}
\frac{1}{\eta^{\hat{1}}}\frac{D^2\eta^{\hat{1}}}{D\tau ^2}=\frac{M_{\star}}{R^3_{\star}}=\kappa \,, \label{roche}
\end{equation}
where $R_{\star}$ and $M_{\star}$ denote the radius and mass of the star, respectively.


In geometrized units, mass and length share the same dimension, and all quantities entering the Roche limit calculation are consistently expressed in units of length. In particular, stellar masses \cite{Misner:1973prb} are converted to geometrized units using $1 M_{\odot}=1.477$km, while stellar radii are given directly in kilometers, as summarized in Table~\ref{tab:stellar_params}

\begin{table}[ht]

\begin{tabular}{lccc}
\hline\hline
Object 
& $M_\star\,[M_\odot]$ 
& $M_\star\,[\mathrm{km}]$ 
& $R_\star\,[\mathrm{km}]$ \\
\hline
Neutron star 
& $1.4$ 
& $2.07$ 
& $12$ \\

Sun-like star 
& $1.0$ 
& $1.477$ 
& $6.96\times10^{5}$ \\


\hline\hline
\end{tabular}
\centering
\caption{Stellar parameters used in the Roche limit calculations. 
Masses are expressed in geometrized units (\(G=c=1\)), while stellar radius are given in km.}
\label{tab:stellar_params}
\end{table}

\subsection{Roche limit for neutron star}

Using \eqref{Fradial} in the canonical sector and inserting it into \eqref{roche}, we can numerically compute the Roche limit $r_{\rm Roche}$ (in geometrized units) for a neutron star orbiting Sgr A* ($M_{\rm Keck}= 3.95\times 10^6 M_{\odot}$ \cite{Do:2019txf}), considering the range of $l$ values constrained by the EHT \cite{Vagnozzi:2022moj}, and keeping $\gamma$ and $Q$ fixed. Specifically: for $l=-4.59\times10^{-3}$, $r_{\rm Roche}=6.71\times 10^5$, with an outer horizon $r_+=1.11707\times 10^7 $. For $l=0$, $r_{\rm Roche}= 6.56\times 10^5$, with $r_+=1.11115\times 10^7$ and for $l=1.24\times 10^{-1}$, the Roche limit $r_{\rm Roche}=3.98\times 10^5$, with horizon $r_+=9.47028\times 10^6$.  Therefore, it can be observed that, for $l>0$ the Roche radius is slightly smaller due to the reduction of the tidal term, while for $l<0$ it increases. 
 
\begin{figure*}[ht!]
\centering
\subfigure[\,\,$Q = 0.3$ and $\gamma=0.3$] 
{\label{Roche1}\includegraphics[width=8.5cm]{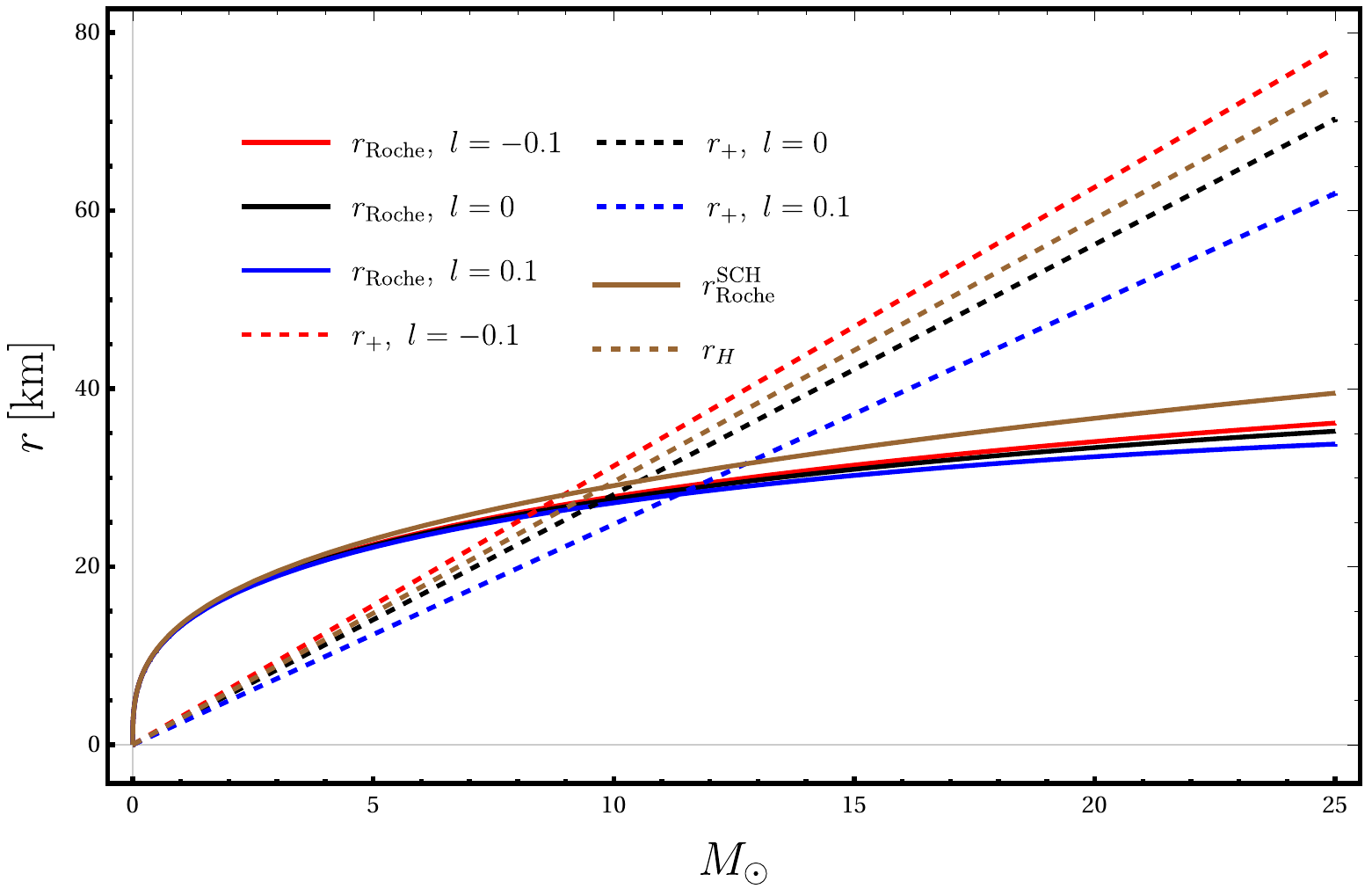} }
\hspace{0.1cm}
\subfigure[\,\, $l=1.24\times 10^{-1}$] 
{\label{Roche2}\includegraphics[width=8.5cm]{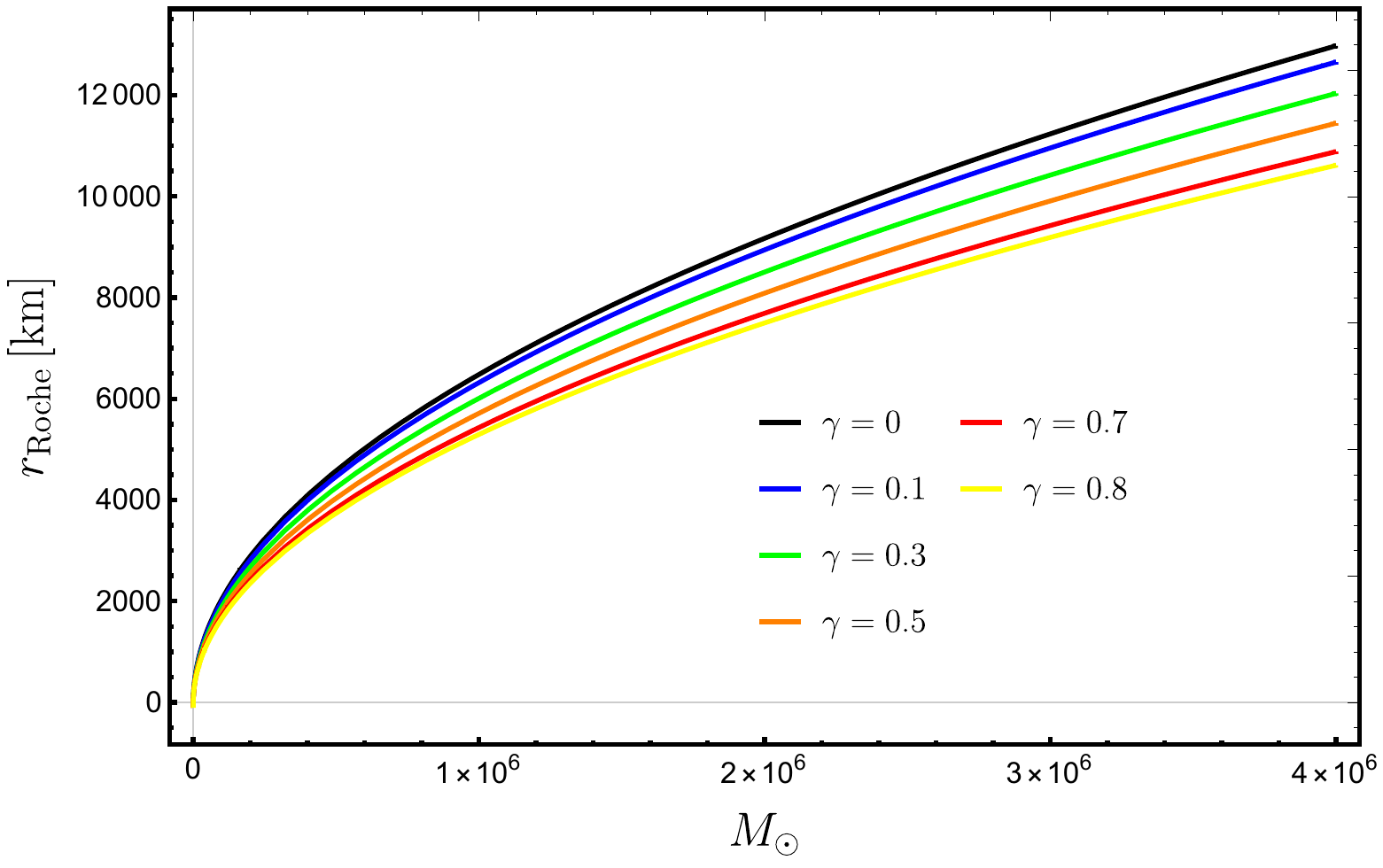} }
\caption{(a) Intersection of the Roche radius (solid curves) for a neutron star profile with the event horizon (dashed curves) for different values of $l$ as a function of the BH mass. (b) Profile of the Roche radius for a neutron star for fixed $l$ and different values of $\gamma$ with $Q=0.3$. Note that we did not used negative values of $l$ because their effects can be neglected within the range obtained in \cite{Duan:2023gng}.}\label{Fig5}
\end{figure*}

For a central object with $\gamma = 0.3$, $Q = 0.5$, and different values of $l$, the intersection between the Roche radius curve and the horizons occurs for masses $M \ll M_{\rm Sgr A^*}$, and is shown in Fig.~\ref{Roche1} for the neutron star case.

The following results can be compared with the Schwarzschild case, for which the Roche limit for a neutron star occurs at $r_{\rm Roche}^{\rm SCH}=r_H=28.9$ for a BH mass of $9.78 M_{\odot}$. This value corresponds to the maximum BH mass for which an observable tidal disruption can occur, commonly referred to as the Schwarzschild Hills mass \cite{Kesden:2011ee, Mummery:2023meb}, and is obtained when the event horizon radius coincides with the Roche radius (intersection between the curve and the brown dotted line). Therefore, the Roche limit is reached when $r_{\rm Roche}=r_+=26.5$ (curve and red dotted line) the BH mass is $8.4M_{\odot}$ for $l=-0.1$. For $l=0$, the Roche limit occurs in $r_{\rm Roche}=r_+ = 27.4$ (curve and black dotted line) when the BH mass is $9.7 M_{\odot}$. For $l=0.1$, the Roche limit occurs in $r_{\rm Roche}=r_+ = 28.1$ (curve and blue dotted line) when the BH mass is $11.3 M_{\odot}$. 

This behavior indicates that, in this parameter regime, tidal disruption takes place before the stellar object reaches the horizon only for relatively low-mass BHs. As the mass increases, the horizon radius grows more rapidly than the Roche radius, preventing tidal disruption outside the horizon. Consequently, for astrophysical supermassive BHs such as Sgr A*, the neutron star would cross the horizon without being tidally disrupted, even in the presence of the KR–ModMax corrections encoded by the parameter $l$.

Figure~\ref{Roche2} illustrates how the ModMax parameter $\gamma$, for fixed $l$, modifies the Roche limit for a neutron star configuration. For $\gamma = 0$ (black curve), the electromagnetic contribution is maximal, which weakens the effective tidal force and requires a larger radius to satisfy the condition given in Eq.~\eqref{roche}. As $\gamma$ increases, the electromagnetic term is progressively suppressed, allowing the gravitational tidal field to dominate at a smaller radius. Consequently, the threshold value $\kappa$ is reached at smaller distances from the central object, leading to a monotonic decrease of the Roche radius with increasing $\gamma$. Although the Roche radius grows with the BH mass and remains ordered for different values of $\gamma$, the event horizon increases more rapidly in the supermassive limit. Consequently, for sufficiently large masses, including that of Sgr A*, the tidal disruption of a neutron star occurs inside the horizon, rendering KR-ModMax corrections phenomenologically irrelevant for observable Roche limit effects.

\subsection{Roche limit for Sun-like star}

\begin{figure*}[ht!]
\centering
\subfigure[\,\,$Q = 0.3$ and $\gamma=0.3$] 
{\label{Roche3}\includegraphics[width=8.5cm]{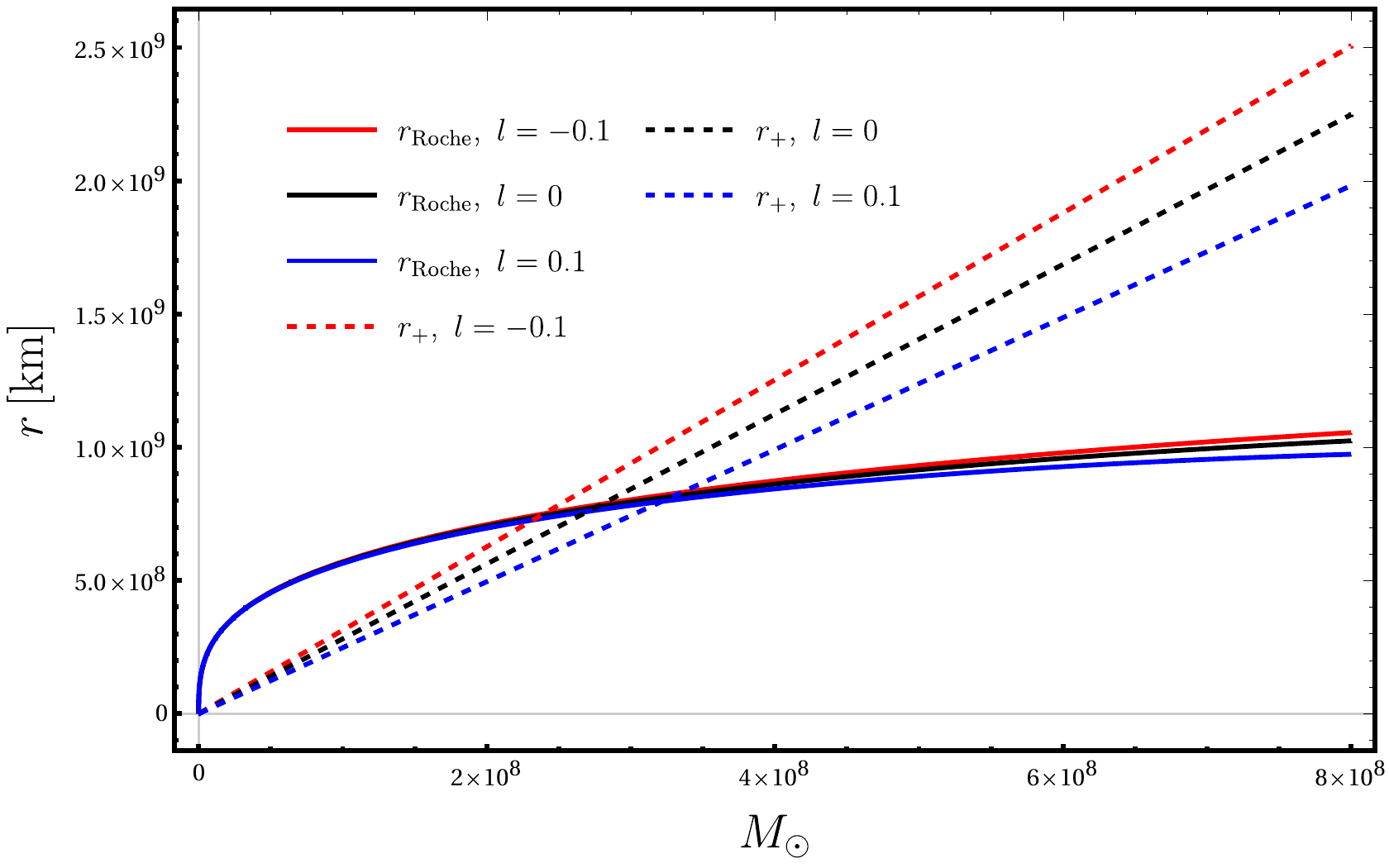} }
\hspace{0.1cm}
\subfigure[\,\, $l=1.24\times 10^{-1}$] 
{\label{Roche4}\includegraphics[width=8.5cm]{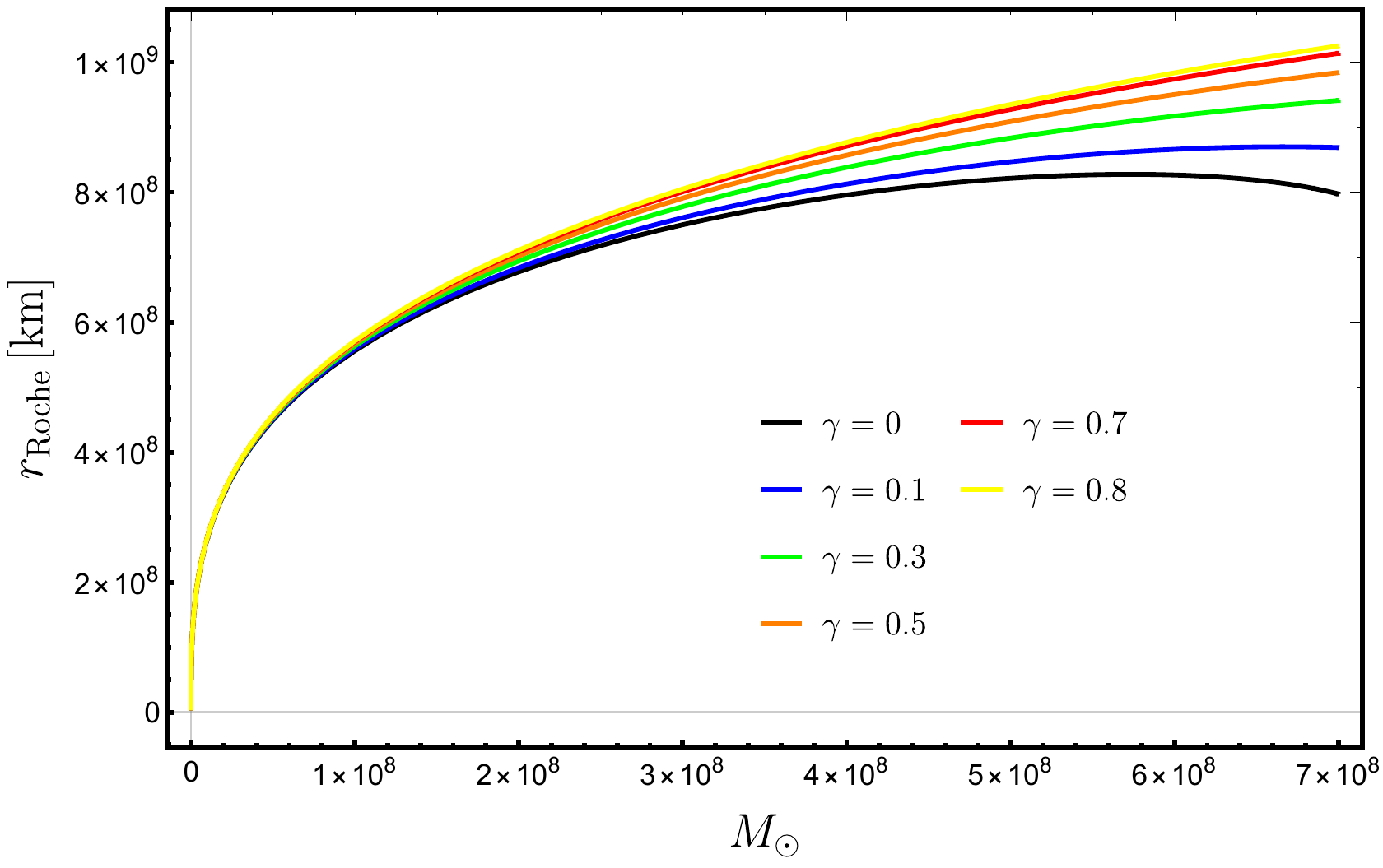} }
\caption{(a) Intersection of the Roche radius (solid curves) for a solar-type profile with the event horizon (dashed curves) for different values of $l$ as a function of the BH mass. (b) Profile of the Roche radius for a solar-type star for fixed $l$ and different values of $\gamma$ with $Q=0.3$. Note that we did not used negative values of $l$ because their effects can be neglected within the range obtained in \cite{Duan:2023gng}.}\label{Fig6}
\end{figure*}

For a Sun-like star with mass $1 M_{\odot}$ approaching a BH described by the KR–ModMax solution, with $\gamma=0.3$, $Q=0.5$, and a mass comparable to that of Sgr A*, the Roche limits for $l=-4.59\times10^{-3}$, $l=0$, and $l=1.24\times10^{-1}$ are nearly identical, with $r_{\rm Roche}\simeq1.98\times10^{8}$. The corresponding horizon radii in this range $l$ are of order $10^{7}$ for $l\leq0$ and $10^{6}$ for $l>0$. The critical configuration defined by $r_{\rm Roche}=r_H$ is shown in Fig.~\ref{Roche3}. Although the dependence on the parameter $l$ follows a trend similar to that found for neutron stars, the tidal disruption of a Sun-like star by a BH with mass comparable to Sgr A* occurs outside the horizon, in contrast to the neutron star case. In this context, the parameter $l$ plays the same qualitative role as in the neutron star analysis, while the stellar compactness governs the distinct disruption outcomes. The corresponding Schwarzschild case for such stars is analogous and follows the same orders of magnitude, with the Schwarzschild Hills mass for a Sun-like star given by $M_{\rm Hills}^{\rm SCH}\simeq 2.8\times10^{8} M_{\odot}$.

Figure~\ref{Roche4} shows how the parameter $\gamma$ modifies the Roche limit as a function of the BH mass. For Sun-like stars, for which the parameter $\kappa$ in Eq.~\eqref{roche} is extremely small, the term proportional to the tidal force $\sim M/r^{3}$ dominates over the electromagnetic contribution. As a consequence, increasing $\gamma$ leads to a reduction of the effective tidal field, requiring a larger radius to satisfy Eq.~\eqref{roche} and thus resulting in an increase of the Roche limit. Notice that the effects of the parameter $\gamma$ become significant only for masses of order $10^{8} M_{\odot}$, which are characteristic of ultramassive BHs \cite{Gezari:2021bmb, Wu:2015phw}.


\section{GEODESIC DEVIATION} \label{secV}
A test particle moving on a timelike geodesic satisfies the condition \cite{dInverno:1992gxs}:
\begin{equation}
g_{\mu\nu}\dot{x}^{\mu}\dot{x}^{\nu}=-1, 
\label{gmn}
\end{equation}
where the overdot denotes differentiation with respect to the affine parameter. In a general static and spherically symmetric metric of the form \eqref{KR0} and considering radial geodesic motion, for which angular momentum is null and $\dot{\theta}= \dot{\phi}=0$ by assumption, then Eq. \eqref{gmn} reads
\begin{equation}
    -A(r)\dot{t}^2+A(r)^{-1}\dot{r}^2=-1\,.
    \label{eq:dsmov}
\end{equation}
In this scenario $E=A(r)\dot{t}$ is the energy of the particle per unit mass, and it is conserved. Replacing this energy conservation in Eq. \eqref{eq:dsmov} we find
\begin{equation}
\frac{dr}{d\tau}=-\sqrt{E^2-A(r)}\,,\label{Eb}
\end{equation}

Let us assume a test particle that moves from rest in the radial position $r=b$. Then its rest energy is given simply by $E=\sqrt{A(r=b)}$. 
Using Eqs.\,\eqref{radial1} and \eqref{angulari}, we obtain
\begin{eqnarray}
\left[E^2-A(r)\right]\frac{d^2\eta^{\hat{1}}}{dr^2}-\frac{A'(r)}{2}\frac{d\eta^{\hat{1}}}{dr}+\frac{A''(r)}{2}\eta^{\hat{1}}&=&0\,,\label{eqdesv1}\\
\left[E^2-A(r)\right]\frac{d^2\eta^{\hat{i}}}{dr^2}-\frac{A'(r)}{2}\frac{d\eta^{\hat{i}}}{dr}+\frac{A'(r)}{2r}\eta^{\hat{i}}&=&0\,.
\label{eqdesvi}
\end{eqnarray}
These equations determine the radial and angular components of the geodesic deviation vectors associated with a freely falling test body in the KR–ModMax BH geometry as functions of the coordinate $r$ \cite{Cordeiro:2025cfo, Crispino:2016pnv}. Solving Eq.~\eqref{eqdesv1} we find
\begin{widetext}
\begin{eqnarray}
\eta^{\hat{1}}&=&\eta^{\hat{1}}(b) \Bigg[ g(r)
- h(r)\sinh ^{-1}\left(\frac{\sqrt{b-r} \sqrt{2 b (1-l)^2 Me^\gamma-\alpha q^2}}{\sqrt{2b}  \sqrt{\alpha q^2-b (1-l)^2 Me^\gamma}}\right)\Bigg]
+\frac{e^\gamma b^3 (1-l)^2}{b (1-l)^2 M e^\gamma - \alpha q^2}\nonumber\\
&&\times \frac{d\eta^{\hat{1}}(b)}{d\tau}
\sqrt{\frac{e^{-\gamma}(r-b) \left(b \left(\alpha q^2-2 (1-l)^2 M e^\gamma r\right)+\alpha q^2 r\right)}{b^2 (1-l)^2 r^2}}\,,\label{Eta1}
\end{eqnarray}
\end{widetext}
where the functions $g(r)$ and $h(r)$ read explicitly to
\begin{eqnarray}
g(r)&=&\frac{1}{2}  \Bigg[3-\frac{r}{b}+\frac{2 \alpha  q^2 (b-r)}{\alpha  q^2 r-b e^{\gamma } (1-l)^2 M r}\nonumber\\
&&+\frac{\alpha  q^2 (b-r)^2}{b r \left(2 b e^{\gamma } (1-l)^2 M-\alpha  q^2\right)}\nonumber\\
&&+\frac{3 \alpha ^2 q^4 (b-r)}{r \left(\alpha  q^2-2 b e^{\gamma} (1-l)^2 M\right)^2}\Bigg]\,,
\end{eqnarray}
and 
\begin{eqnarray}
h(r)&=&\frac{6 e^\gamma b^{3/2} (1-l)^2 M\sqrt{b-r}}{r \left( 2  b (1-l)^2 M e^\gamma- \alpha q^2\right)^{5/2}}\nonumber\\
&&\times\left(\alpha q^2-b (1-l)^2 M e^\gamma \right)^{1/2} \nonumber\\
&&\times\sqrt{\left(\alpha q^2-2 e^\gamma (1-l)^2 M r\right)+\alpha q^2 r/b}\,. 
\end{eqnarray}
We also have that $\eta^{\hat{1}}(b)$ and $\frac{d\eta^{\hat{1}}(b)}{d\tau}$ are integration constants which will be determined by the appropriate initial conditions.

The angular component reads as
\begin{eqnarray}
&&\eta^{\hat{i}}=
r \Bigg[\frac{1}{b}\eta^{\hat{i}}(b)+\frac{2 e^{\gamma/2} b  (1-l)}{q\sqrt{\alpha }}\frac{d\eta^{\hat{i}}(b)}{d\tau}\nonumber\\
&&\times\tan ^{-1}\left(\frac{q \sqrt{\alpha(b-r)}}{\sqrt{2 b e^{\gamma } (l-1)^2 M r-\alpha  q^2 (b+r)}}\right)\Bigg],
\end{eqnarray}
with $\eta^{\hat{i}}(b)$ and $\frac{d\eta^{\hat{i}}(b)}{d\tau}$ being integration constants. 

Here, it is important to note the presence of the term that controls the canonical and phantom sectors. In the phantom sector, the angular geodesic deviation admits a real representation in terms of hyperbolic functions,
\begin{eqnarray}
    &&\eta^{\hat{i}}=r \Bigg[\frac{1}{b}\eta^{\hat{i}}(b)+\frac{2 e^{\gamma/2} b  (1-l)}{q}\frac{d\eta^{\hat{i}}(b)}{d\tau}\nonumber\\ &&\times\tanh ^{-1}\left(\frac{q \sqrt{(b-r)}}{\sqrt{2 b e^{\gamma } (l-1)^2 M r+ q^2 (b+r)}}\right)\Bigg].
\end{eqnarray}

\begin{figure*}[t!]
\centering
\subfigure[\,\,$Q = 0.6 $ and $\gamma=0.5$ ] 
{\label{Figdesva2}\includegraphics[width=8.5cm]{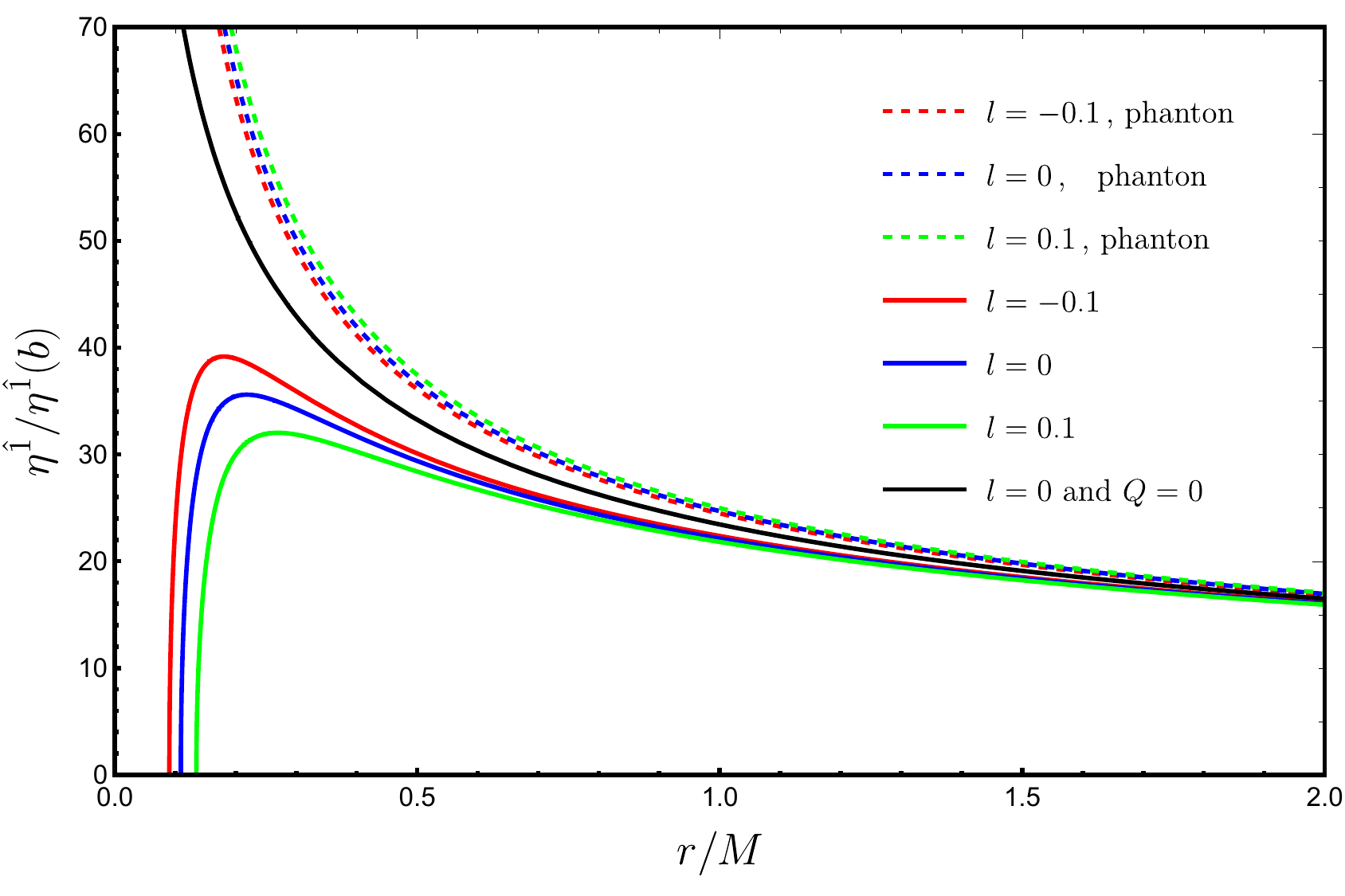} }
\hspace{0.1cm}
\subfigure[\,\,$l=1.24\times 10^{-1}$ ] 
{\label{Figdesvb2}\includegraphics[width=8.5cm]{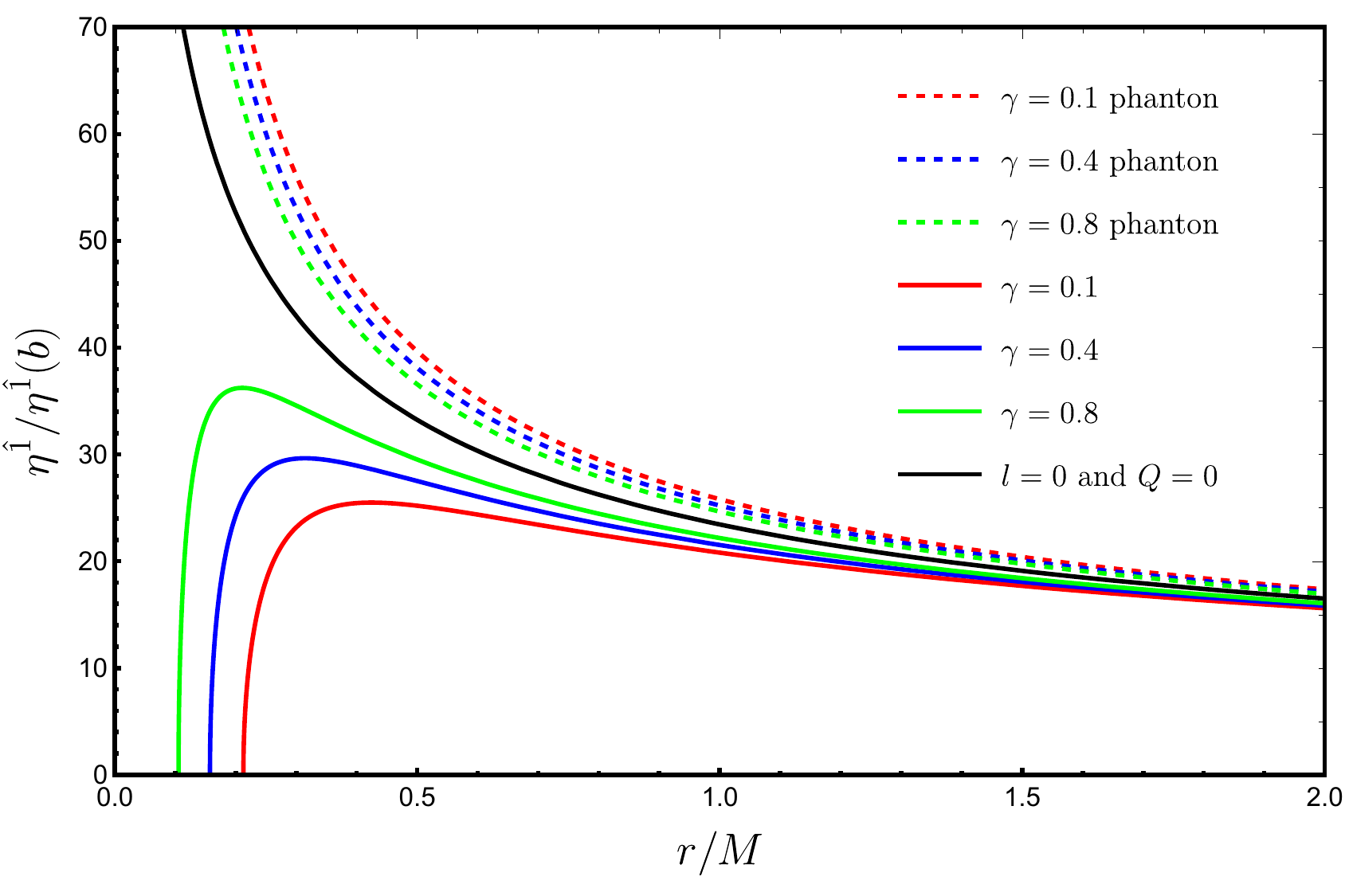} }
\caption{In (a) we have plotted the radial component of the geodesic deviation for different values of $l$ with fixed $Q$ and $\gamma$. In (b) we have plotted radial geodesic deviation for different values of $\gamma$ and fixed $l$ with $Q=0.6$. Both of them for condition $\mathcal{CI}$.}\label{Fig8}
\end{figure*}

In our analysis of the geodesic deviation, we consider two sets of initial conditions. First, a test body is initially at rest at $r=b$ and falls toward the BH of the KR-ModMax solution, subject to $\mathcal{CI}=\lbrace\eta^{\hat{1}}(b)>0\,,\, d\eta^{\hat{1}}(b)/d\tau=0\,, \, \eta^{\hat{i}}>0\,,\, d\eta^{\hat{i}}(b)/d\tau=0\rbrace$. Second, the test body is now subject to the condition $\mathcal{CII}=\lbrace\eta^{\hat{1}}(b)=0\,,\, d\eta^{\hat{1}}(b)/d\tau>0\,, \, \eta^{\hat{i}}(b)=0\,,\, d\eta^{\hat{i}}(b)/d\tau>0\rbrace$, which represents a body exploding from a point at $r=b$. We graphically analyze the geodesic deviation of the KR-ModMax solution in both the canonical sector ($\alpha=1$) and the phantom sector ($\alpha=-1$), with respect to the LV parameter $l$ and the ModMax parameter $\gamma$, for a fixed value $b=100$. For condition $\mathcal{CI}$, the radial components of the geodesic deviation are shown in Fig.\,\ref{Fig8}. 

In the canonical sector, for different values of $l$, Fig.\,\ref{Figdesva2} shows that as the particle approaches the outer horizon, the radial component \eqref{Eta1} increases until it reaches a maximum limiting value and then decreases toward the inner horizon $r_{-}$, eventually vanishing. The maximum value of this component decreases as $l$ increases from negative values. For fixed $l>0$  Fig.\,\ref{Figdesvb2} shows that increasing the ModMax parameter $\gamma$ leads to a higher maximum in the radial geodesic deviation, preserving the qualitative profile observed in Fig.\,\ref{Figdesva2}. The results suggest that the ModMax parameter $\gamma$ strengthens the tidal response of the spacetime, intensifying the radial geodesic deviation while leaving its qualitative behavior unchanged.

This behavior is similar to the RN case reported in \cite{Crispino:2016pnv} and to the charged KR solution discussed in \cite{Cordeiro:2025cfo}, where there exist regions in which the radial component is compressed rather than stretched. However, it is opposite to what is observed in the Schwarzschild case (black curve) and in the phantom sector of the KR–ModMax solution (dotted curves), where, as the body crosses the event horizon, the radial tidal force diverges. 

For the chosen initial condition, the angular component of the geodesic deviation reduces to $\eta^{\hat{i}}=\frac{r}{b}\eta^{\hat{i}}(b)$, exactly the same form as in the Schwarzschild, RN, and charged KR cases \cite{Crispino:2016pnv, Cordeiro:2025cfo}, which is a direct consequence of spherical symmetry and of the chosen initial condition, rather than of the specific matter content. As a result, the angular deviation vanishes in the center and increases linearly with the radial coordinate. 

\begin{figure*}[t!]
\centering
\subfigure[\,\,$Q = 0.6$ and $\gamma=0.5$] 
{\label{Figdesva3}\includegraphics[width=8.5cm]{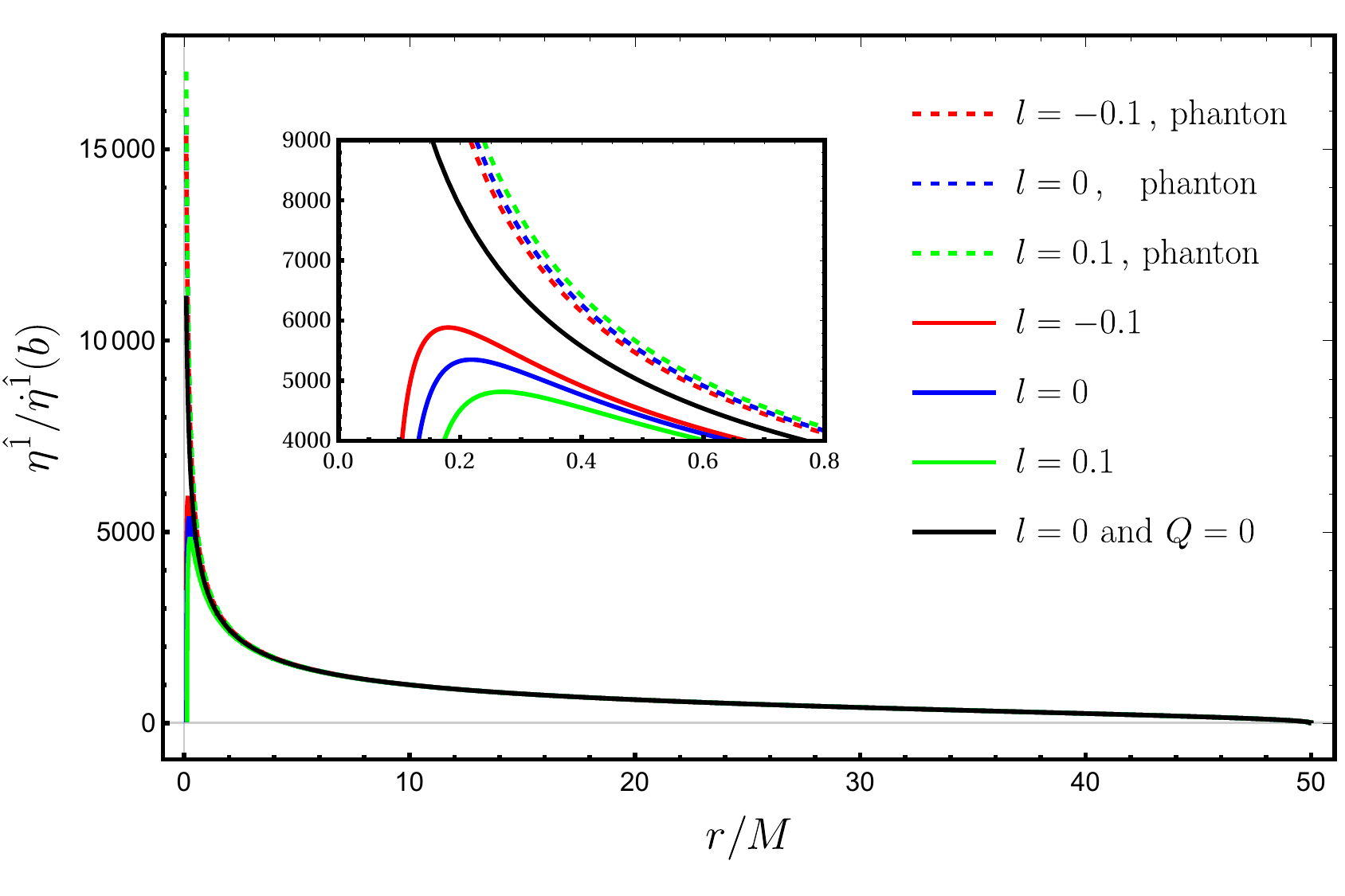} }
\hspace{0.1cm}
\subfigure[\,\,$l=1.24\times 10^{-1}$ ] 
{\label{Figdesvb3}\includegraphics[width=8.5cm]{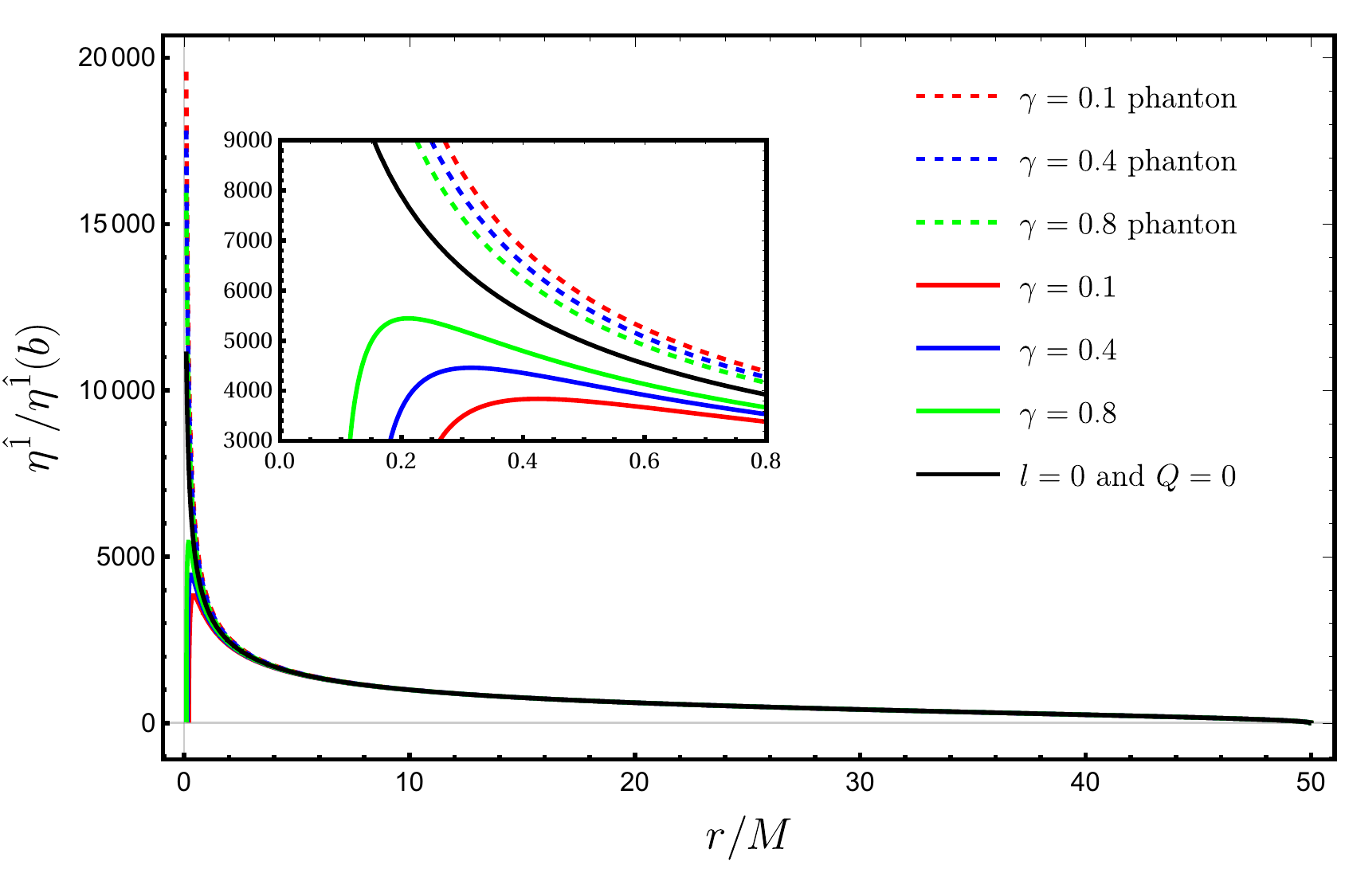} }
\caption{In (a) we have plotted the radial component of the geodesic deviation for different values of $l$ with fixed $Q$ and $\gamma$. In (b) we have plotted radial geodesic deviation for different values of $\gamma$ and fixed $l$ with $Q=0.6$. Both of them for condition $\mathcal{CII}$.}\label{Fig9}
\end{figure*}

\begin{figure*}[t!]
\centering
\subfigure[\,\,$Q = 0.6 $ and $\gamma=0.5$ ] 
{\label{Figdesva4}\includegraphics[width=8.5cm]{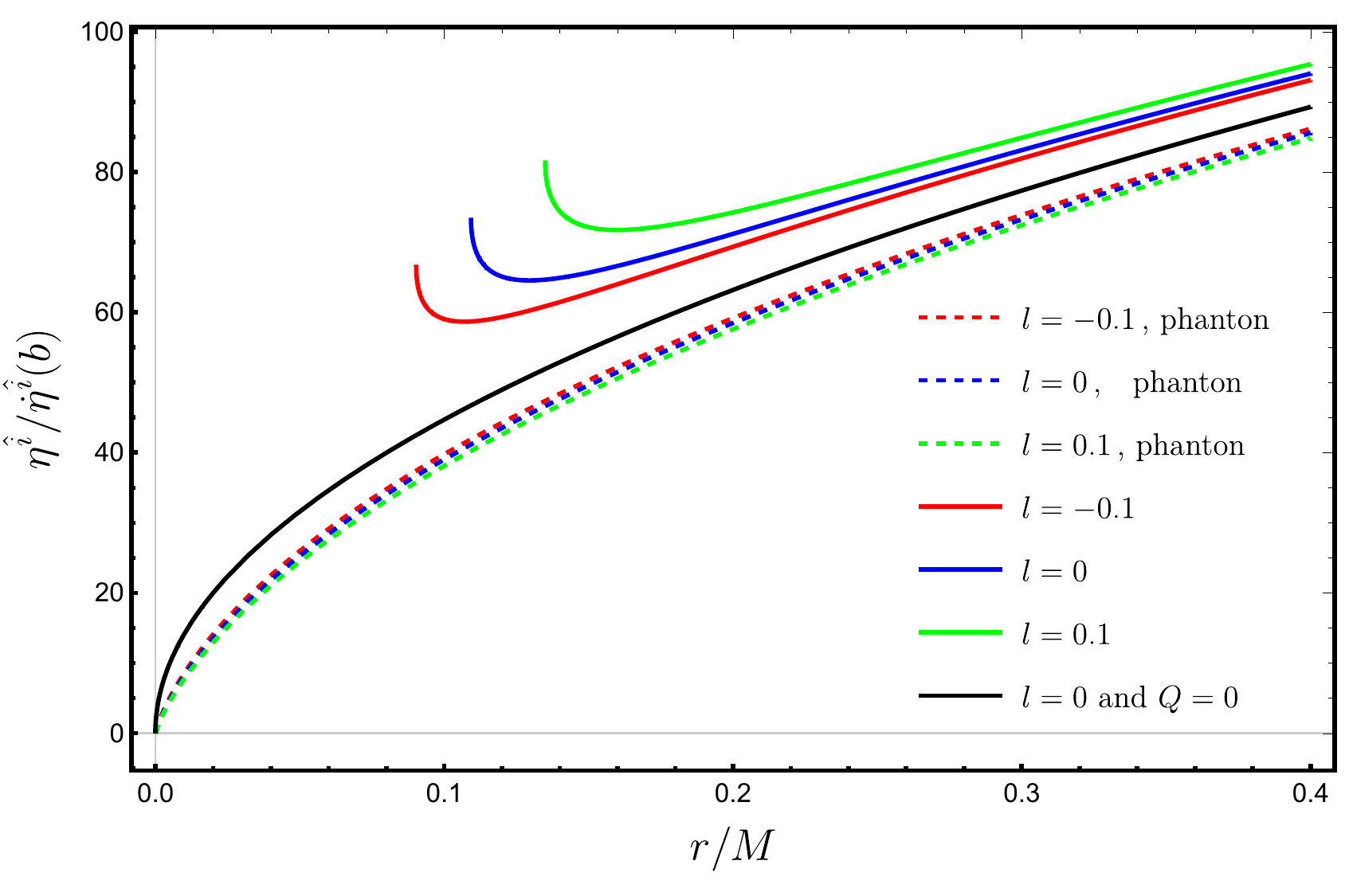} }
\hspace{0.1cm}
\subfigure[\,\,$l=1.24\times 10^{-1}$ ] 
{\label{Figdesvb4}\includegraphics[width=8.5cm]{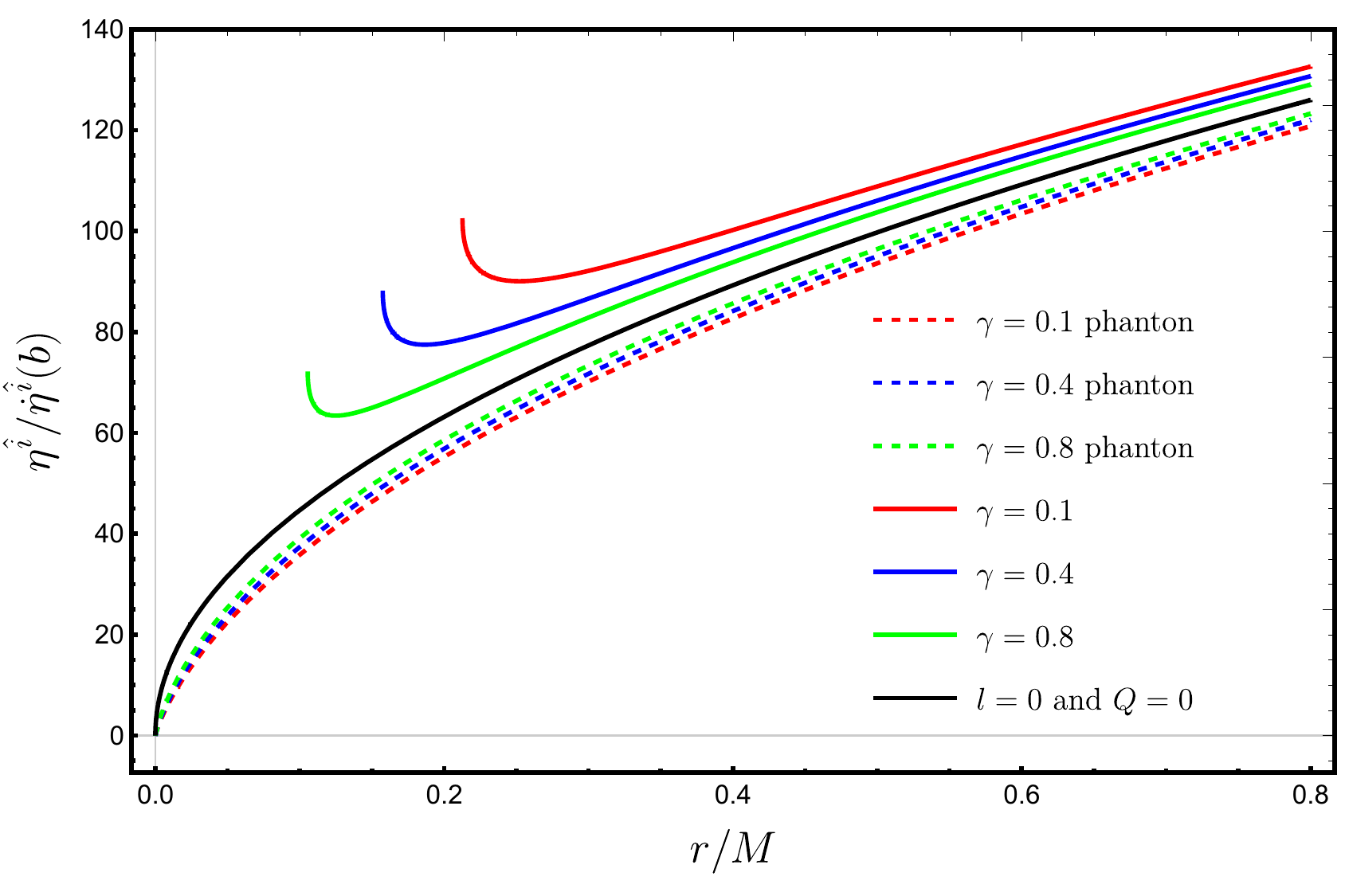} }
\caption{In (a) we have plotted the angular component of the geodesic deviation for different values of $l$ with fixed $Q$ and $\gamma$. In (b) we have plotted radial geodesic deviation for different values of $\gamma$ and fixed $l$. Both of them for condition $\mathcal{CII}$.}\label{Fig10}
\end{figure*}

Figure\,\ref{Fig9} shows the radial component of $\mathcal{CII}$, which exhibits a qualitatively similar behavior to that observed for condition $\mathcal{CI}$, both in the canonical and phantom sectors. For both initial conditions $\mathcal{CI}$ and $\mathcal{CII}$, the divergent behavior observed in the phantom sector reflects a genuine tidal instability associated with the violation of the energy condition \cite{Sekhmani:2025epe}. The LV parameter $l$ affects this sector only subtly, modifying the strength of the instability while preserving its overall qualitative structure.
The angular component is shown in Fig.\,\ref{Fig10}.  
In the canonical sector, for a body released at $b=100$, the angular component of the geodesic deviation increases until it reaches a maximum at $r=b/2$, after which it begins to decrease due to the onset of tidal compressive forces. The influence of the LV parameter $l$ (Fig.\,\ref{Figdesva4}) is mainly confined to the interior region, becoming significant only in the vicinity of $r=0$. In this regime, negative values of $l$ lead to a smaller angular geodesic deviation, whereas positive values enhance the deviation when compared with the reference case $l=0$. As the body moves further inward, the angular deviation reaches a minimum and subsequently increases, eventually attaining a stopping point at $r=R^{\rm stop}$. Fig.\,\ref{Figdesvb4} illustrates the role of the ModMax parameter $\gamma$ in the KR-ModMax solution. Although the qualitative structure of the angular geodesic deviation remains similar to that observed for different values of $l$, the physical effect of $\gamma$ is fundamentally different. Increasing $\gamma$ suppresses the magnitude of the angular deviation, shifting the stopping point $r=R^{\rm stop}$ towards smaller values of $r$. This behavior indicates that, while both parameters $l$ and $\gamma$ primarily affect the strong field regime near $r=0$, they do so through different mechanisms: $l$ introduces localized geometric modifications associated with LV effects, whereas $\gamma$ encodes nonlinear electromagnetic contributions that regulate the tidal response without altering the overall qualitative structure. In the phantom sector, the presence of the ModMax parameter $\gamma$ and the LV parameter $l$ does not lead to qualitative modifications of the tidal structure, but rather results in a mild attenuation of the deviation amplitude. Consequently, the overall Schwarzschild-like profile of the angular geodesic deviation is preserved, with $\gamma$ and $l$ acting only as secondary regulators of the tidal response.

\section{Summary and conclusion}\label{Sec:Conclusion}
In this work we analyze tidal forces, geodesic deviation, and the Roche limit in the BH spacetime described by the KR-ModMax solution \eqref{KR0}, in which electromagnetic nonlinearity is controlled by the parameter $\gamma$ and LV by the parameter $l$. We show that whenever the charge $Q\neq0$, these parameters introduce both qualitative and quantitative modifications to the tidal structure, distinguishing the KR-ModMax system from the Schwarzschild, RN, and charged KR scenarios.

In Sec.~\ref{sec:two}, we show that, in the canonical sector ($\alpha=1$), the radial tidal force exhibits a transition region characterized by a sign inversion at $r=R^{\rm rad}_{0}$, located between the inner and outer horizons $r_{-}$ and $r_+$. This structure indicates the existence of internal regimes in which extended objects cease to be radially stretched and instead undergo compression, in analogy with charged BHs. The LV parameter $l$ acts directly on the intensity and location of these regions, amplifying tidal effects for $l<0$ and attenuating them for $l>0$ through modifications of the fundamental geometric scales of spacetime. This behavior closely resembles that found in \cite{Crispino:2016pnv} for the RN case and in \cite{Cordeiro:2025cfo} for the charged KR solution with Maxwell coupling. In turn, the parameter $\gamma$ regulates the nonlinear contribution of the electromagnetic sector, strengthening the radial tidal response as it increases, without altering the qualitative structure of the transition.

The angular component of the tidal force in the canonical sector is predominantly compressive, in agreement with the RN and charged KR cases discussed above. In this context, $l$ controls the effective geometry and the location of the extrema of the angular tidal force, while $\gamma$ acts as a damping factor, progressively suppressing the nonlinear ModMax effects through the term $e^{-\gamma}Q^2$. In this way, the parameters $l$ and $\gamma$ play complementary roles: the former shapes the geometric structure of spacetime, whereas the latter regulates the magnitude and smoothness of the tidal response.

In the phantom sector ($\alpha=-1$), both the tidal forces and the geodesic deviation exhibit divergences associated with the violation of the energy conditions, characterizing a genuine tidal instability. In this regime, the parameters $l$ and $\gamma$ affect only the strength of the forces, preserving a global structure qualitatively similar to that of the Schwarzschild case. In particular, the angular component is qualitatively insensitive to variations of $\gamma$, reinforcing the universal character of the tidal instability in the phantom sector.

In Sec.~\ref{secIV}, the analysis of the Roche limit shows that the parameter $l$ directly modifies the hierarchy between the horizon radius $r_+$ and the tidal disruption radius $r_{\rm Roche}$, shifting the critical mass (Hills mass) defined by the condition $r_{\rm Roche}=r_+$. For neutron stars, even when considering values $l$ compatible with the observational constraints from the EHT, tidal disruption occurs inside the horizon for supermassive BHs such as Sgr A*, rendering KR-ModMax corrections observationally irrelevant. In contrast, for Sun-like stars, tidal disruption takes place outside the horizon, preserving the possibility of observable tidal disruption events. In this case, the parameter $l$ systematically shifts the critical mass, while the effects of $\gamma$ become relevant only for ultramassive BHs with masses of order $10^{8}M_{\odot}$ \cite{Gezari:2021bmb, Wu:2015phw}.

Finally, in Sec.~\ref{secV}, the geodesic deviation analysis confirms that, in the canonical sector, the parameters $l$ and $\gamma$ mainly affect the magnitude of the tidal response in the strong-field regime, without modifying its qualitative structure. The parameter $l$ introduces localized geometric modifications, whereas it $\gamma$ regulates the intensity of the response associated with electromagnetic nonlinearity, shifting characteristic points such as the stopping radius $r=R^{\rm stop}$. In the phantom sector, the persistent divergence of the geodesic deviation reinforces the interpretation of an intrinsic tidal instability, only mildly modulated by $l$ and $\gamma$.

Taken together, our results show that the KR-ModMax system provides a consistent theoretical laboratory to investigate how LV and NED affect tidal structures and stellar disruption processes in strong-field regimes. Although such effects are largely hidden for supermassive BHs, they may become relevant for intermediate-mass BHs and for tidal disruption events involving low-compactness stars, offering a possible, albeit indirect, way to test extensions of standard GR.

\section*{Acknowledgements}

M.S. thanks Conselho Nacional de Desenvolvimento Cient\'ifico e Tecnol\'ogico - CNPq, Brazil, CNPQ/PDE 200218/2025-5, for financial support.

\bibliography{ref.bib}

\end{document}